\documentclass[fleqn,usenatbib]{mnras}

\usepackage{newtxtext}

\usepackage[TU]{fontenc}

\usepackage{ae,aecompl}

\usepackage[utf8]{inputenc}
\DeclareUnicodeCharacter{2212}{-}

\usepackage{graphicx}	
\usepackage{amsmath}	
\usepackage{amssymb}	
\usepackage[british]{babel}             
\usepackage{bm}
\usepackage{physics}
\usepackage[dvipsnames]{xcolor}
\definecolor{orcidlogocol}{HTML}{A6CE39}
\usepackage{booktabs}
\usepackage{makecell}
\usepackage{ulem}

\usepackage[nolist]{acronym}
\newacro{cpd}[CPD]{circumplanetary disk}
\newacro{RSDs}{redshift space distortions}
\newacro{RSD}{redshift space distortions}
\newacro{DGP}{Dvali-Gabadadze-Porrati}
\newacro{CMB}{cosmic microwave background}
\newacro{SM}{Streaming Model}
\newacro{GSM}{Gaussian Streaming Model}
\newacro{STSM}{Skew-T Streaming Model}

\newacro{LPT}{Lagrangian Perturbation Theory}
\newacro{CLPT}{Convolutional Lagrangian Perturbation Theory}
\newacro{GR}{General Relativity}
\newacro{MG}{Modified Gravity}
\newacro{ST}{skewed student-t}
\newacro{PDF}{Probability Distribution Function}
\newacro{FFT}{Fast Fourier Transform}
\newacro{HOD}{Halo Occupation Distribution}
\newacro{SHAM}{subhalo abundance matching}
\newacro{HMF}{halo mass function}
\newacro{LSS}{large-scale structure}
\newacro{TPCF}{two-point correlation function}
\newacro{EFT}{effective field theory}

\title[Small-scale RSDs in MG theories]{Towards an accurate model of small-scale redshift-space distortions in modified gravity}

\author[C. Ruan et al.]{Cheng-Zong Ruan$^{1}$\thanks{E-mail: cheng-zong.ruan@durham.ac.uk},
Carolina Cuesta-Lazaro$^{1,2}$,
Alexander Eggemeier$^{1}$,\
\newauthor C\'{e}sar Hern\'{a}ndez-Aguayo$^{3,4}$,
Carlton M. Baugh$^{1,2}$, Baojiu Li$^{1}$ and Francisco Prada$^{5}$ \\
$^{1}$Institute for Computational Cosmology, Department of Physics, Durham University, South Road, Durham DH1 3LE, UK\\
$^{2}$Institute for Data Science, Durham University, South Road, Durham DH1 3LE, UK\\
$^{3}$Max-Planck-Institut f\"ur Astrophysik, Karl-Schwarzschild-Str 1, D-85748 Garching, Germany\\
$^{4}$Excellence Cluster ORIGINS, Boltzmannstrasse 2, D-85748 Garching, Germany \\
$^{5}$Instituto de Astrof\'{i}sica de Andaluc\'{i}a (CSIC), Glorieta de la Astronom\'{i}a, E-18080 Granada, Spain
}

\date{Accepted XXX. Received YYY; in original form \today}

\pubyear{2021}

\begin{document}
\label{firstpage}
\pagerange{\pageref{firstpage}--\pageref{lastpage}}
\maketitle

\begin{abstract}
    The coming generation of galaxy surveys will provide measurements of galaxy clustering with unprecedented accuracy and data size, which will allow us to test cosmological models at much higher precision than achievable previously. This means that we must have more accurate theoretical predictions to compare with future observational data.
    As a first step towards more accurate modelling of the
    redshift space distortions (RSD) of small-scale galaxy clustering in modified gravity (MG) cosmologies, we investigate the validity of the so-called Skew-T (ST) probability distribution function (PDF) of halo pairwise peculiar velocities in these models.
    We show that, combined with the streaming model of RSD, the ST PDF substantially improves the small-scale predictions by incorporating skewness and kurtosis, for both $\Lambda$CDM and two leading MG models: $f(R)$ gravity and the DGP braneworld model.
    The ST model reproduces the velocity PDF and redshift-space halo clustering measured from MG $N$-body simulations very well down to $\sim 5 \, h^{-1}\mathrm{Mpc}$. In particular, we investigate the enhancements of halo pairwise velocity moments with respect to $\Lambda$CDM for a larger range of MG variants than previous works, and present simple explanations to the behaviours observed. By performing a simple Fisher analysis, we find a significnat increase in constraining power to detect modifications of General Relativity by introducing small-scale information in the RSD analyses.
\end{abstract}




\section{Introduction}
\label{sec:intro}

The standard $\Lambda$ cold dark matter ($\Lambda$CDM) cosmological model, in which gravity is described by Einstein's theory of  \ac{GR}, provides an excellent fit to various cosmological observations, such as the cosmic microwave background anisotropies \citep[e.g.][]{Planck15Overview:2016A&A...594A...1P}, weak and strong gravitational lensing \cite[e.g.][]{Kilbinger:2015RPPh...78h6901K,DESY3CosConstraints:2021arXiv210513546P,H0licowOverview:2017MNRAS.468.2590S}, and the large-scale clustering of galaxies \citep[e.g.][]{Alam:2017MNRAS.470.2617A,eBOSSOverview:2021PhRvD.103h3533A}. However, the nature of the cosmological constant ($\Lambda$) still lacks a reliable physical explanation, as the vacuum energy density predicted by the standard model of particle physics is many orders of magnitude larger than the value of $\Lambda$ inferred from cosmological observations \citep{CarrollLambda:2001LRR.....4....1C}. Instead of invoking a finely tuned $\Lambda$ or other exotic dark energy components, alternative approaches assume that \ac{GR} might be inaccurate on cosmic scales, and that \ac{MG} scenarios are plaussible alternatives to the standard laws of gravity \citep[see e.g.][for recent reviews]{Joyce:2015PhR...568....1J,Koyama:2018IJMPD..2748001K,2019ARA&A..57..335F}. Some leading examples of \ac{MG} theories include the \ac{DGP} braneworld model \citep{DvaliDGP:2000PhLB..485..208D}, the symmetron \citep{Hinterbichler:2010PhRvL.104w1301H,Hinterbichler:2011PhRvD..84j3521H}, the k-mouflage model \citep{Babichev:2009IJMPD..18.2147B}, and $f(R)$ gravity \citep{Sotiriou:2010RvMP...82..451S,DeFelice:2010LRR....13....3D} which is a particular subclass of the so-called chameleon models \citep{Khoury:2004PhRvL..93q1104K,Khoury:2004PhRvD..69d4026K,Brax:2008PhRvD..78j4021B}. By considering different \ac{MG} models we can see how alternatives to \ac{GR} might change measurable quantities, and therefore suggest which observables have the most potential to constrain deviations from \ac{GR}.

Modifications to \ac{GR} typically manifest themselves as changes to the cosmic expansion history and/or the evolution of structure, i.e., at the background and/or perturbation levels.
Many viable \ac{MG} models closely mimic the expansion history of $\Lambda$CDM, and therefore are hard to distinguish from \ac{GR} using background cosmology alone. Even in the case of \ac{MG} models where the expansion rate is modified, there can be degeneracies which can not be fully broken using background observables. Hence, high hopes have been placed on the use of cosmological observations that involve perturbation dynamics to test gravity. The evolution of perturbations in linear theory in \ac{MG} models has been well-studied both theoretically \citep[e.g.][]{Brax:2011PhRvD..84l3524B,Barreira:2012PhRvD..86l4016B,Barreira:2015PhRvD..91f3528B} and numerically \citep[e.g.][]{HojjatiMGCAMB:2011JCAP...08..005H,HuEFTCAMB:2014PhRvD..89j3530H,Bellini:2018PhRvD..97b3520B}, and the model predictions have been confronted with observational data such as the \ac{CMB} temperature fluctuations and the matter power spectrum \citep{Hu:2013PhRvD..88l3514H,Dossett:2014JCAP...03..046D}. Nevertheless, on small scales where tremendous amounts of observational data are available, linear theory breaks down and a fully non-linear treatment is needed in order to more accurately predict the model behaviour. An improved non-linear model is essential if, for example, one wishes to make the best use of the current and next generations cosmological surveys to test models. This point becomes even more acute in the context of \ac{MG} cosmology, given that such models have intrinsically non-linear features, such as screening mechanisms, which cannot be captured by linear theory \citep[e.g.][]{Li:2013MNRAS.428..743L}. For this reason, here we focus on non-linear structure formation in \ac{MG} cosmologies, with the objective of improving model tests by including data from the non-linear regime of the \ac{LSS} of the Universe.

One of the most commonly-used probes of the \ac{LSS} is galaxy clustering, which records the angular positions and redshifts (as proxies for radial distance) of galaxies. The measured redshifts of galaxies are affected by their peculiar velocities, which cause an anisotropy in the estimated galaxy clustering (in redshift-space)---known as \ac{RSD}---since the line-of-sight direction of the observer is singled out as being special, and the peculiar motions cannot be separated from the Hubble expansion in this direction. \ac{RSD} encode cosmological information about both the spatial distribution and the velocity field of galaxies, which makes them a useful probe of the laws of gravity \citep[e.g.,][]{Kaiser:1987MNRAS.227....1K,Hamilton:1992ApJ...385L...5H,Guzzo:2008Natur.451..541G,Song:2015PhRvD..92d3522S,Barreira:2016PhRvD..94h4022B,He:2018oai}, because peculiar velocities are mainly induced by the gravity of the inhomogeneous matter distribution.

Current best  constraints on the growth rate of the \ac{LSS} from the  \ac{TPCF} in redshift-space are consistent with \ac{GR} (see, e.g., \citealt{Bautista_eBOSSLRGConfigSpace:2021MNRAS.500..736B} for an analysis using the luminous red galaxy sample from eBOSS and \citealt{Hou_eBOSSQSOConfigSpace:2021MNRAS.500.1201H} for another using eBOSS quasars). 
Various perturbation theory-based methods have been used to model RSD: these include the combined \ac{GSM} and \ac{CLPT} formalism developed by \citet{Reid:2011MNRAS.417.1913R,Carlson:2013MNRAS.429.1674C,Wang:2014MNRAS.437..588W}, models such as those described in  \citet{Taruya_TNS:2010PhRvD..82f3522T} (TNS) and \citet{2017MNRAS.464.1640S}, which derive from the perturbative expansion advocated in \citet{2004PhRvD..70h3007S}, as well as effective field theory approaches (e.g., \citealt{2014arXiv1409.1225S,2020JCAP...07..011F}). However, the validity of these approaches is typically limited to sufficiently large scales, where perturbations can be considered linear or quasi-linear. For example, for the eBOSS luminous red galaxy sample \citet{Bautista_eBOSSLRGConfigSpace:2021MNRAS.500..736B} found that, to achieve unbiased constraints on the cosmological parameters with the CLPT model, the minimun scale to be used in the fitting process is $20 \, h^{-1} \mathrm{Mpc}$.
Current and upcoming spectroscopic redshift measurements, such as DESI \citep{DESI:2016arXiv161100036D} and Euclid \citep[][]{Euclid:2011arXiv1110.3193L,AmendolaEuclid:2013LRR....16....6A}, will provide galaxy power spectrum and correlation function measurements with much higher accuracy than currently available and down to smaller scales, which places a much stronger demand on the accuracy of \ac{RSD} modelling, if we are to fully exploit these observational clustering estimates.

Here, to tackle this challenge, we adopt the \ac{SM}, which was introduced by \citet{Peebles:1980lssu.book.....P} and further investigated by \citet{1995ApJ...448..494F,2004PhRvD..70h3007S}, and which is widely used nowadays, to model the redshift-space \ac{TPCF}s in \ac{MG} models. The streaming model takes the real-space two-point correlation function and the galaxy pairwise velocity \ac{PDF} as ingredients (cf.~Eq.~\eqref{eqn:streaming_model_core} below).
The former is related to the matter clustering in real space, and the latter encodes the physics underlying the evolution of peculiar velocities induced by gravitational instability.
Instead of the usual \ac{GSM}, we will follow  \citet{Cuesta-Lazaro:2020MNRAS.498.1175C} who showed that a one-dimensional \ac{ST} distribution can accurately describe the \ac{PDF} of the line-of-sight pairwise velocity for dark matter haloes down to small scales. This model has been validated against simulations of the $\Lambda$CDM cosmology, in terms of both the velocity \ac{PDF} itself and the predictions of correlation function multipoles. 

We extend the evaluation of the non-linear \ac{RSD} model carried out by \citet{Cuesta-Lazaro:2020MNRAS.498.1175C} to \ac{MG} cosmologies. Given the ever declining sample variance errors expected from upcoming large-scale structure measurements, and the small differences expected between the predictions of viable gravity models, it is imperative to produce accurate models of RSD in different cosmologies. 
Thanks to significant recent progress, modern \ac{MG} $N$-body codes are now capable of running large-volume and high-resolution simulations to meet the requirements of upcoming wide field galaxy surveys. We test the validity of the \ac{ST} distribution using $N$-body simulations, based on the newly-developed \ac{MG} code \textsc{mg-glam} \citep{Hernandez-Aguayo:MG-GLAM2021,Ruan:2021MGGLAMfR}, which enables the fast generation of simulations in a wide range of \ac{MG} models. 

Our aim is to investigate if the \ac{ST} pairwise velocity \ac{PDF} for dark matter halos, a generic phenomenological model that is applicable to a wide range of \ac{MG} cosmologies, at a similar level of accuracy  as for the $\Lambda$CDM model \citep{Cuesta-Lazaro:2020MNRAS.498.1175C}. We find that the streaming model of \ac{RSD} combined with the \ac{ST} velocity \ac{PDF} reproduces the halo clustering multipoles measured from $N$-body simulations down to $\simeq 5 \, h^{-1}\mathrm{Mpc}$ for all gravity models considered.
We also explore the behaviour of halo pairwise velocity moments in two representative classes of \ac{MG} models, $f(R)$ gravity and the normal branch of \ac{DGP} gravity, along with their relative differences with respect to $\Lambda$CDM. Finally, we show that including small-scale RSD can indeed lead to greatly improved constraints on these models.

This paper is organised as follows. In Section~\ref{sec:simulations}, we give a brief description of 
the \ac{MG} models considered and the $N$-body simulations used in our analysis. In Section~\ref{sec:STmodel}, we review the streaming model of \ac{RSD}, with a particular focus on one of its ingredients --- the pairwise velocity \ac{PDF} of dark matter halos. In Section~\ref{sec:res}, we study the behaviour of the halo pairwise velocity \ac{PDF}s in a wide range of \ac{MG} models, show that the streaming model with the \ac{ST} \ac{PDF} accurately reproduces redshift-space two-point correlation functions, and perform a simple Fisher matrix analysis to assess the impact of including small-scale RSD on the model constraints. Finally, we summarise and conclude in Section~\ref{sec:conc_and_disc}. Throughout, our analysis is based on dark matter haloes, and we leave the extension of the \ac{RSD} modelling to galaxy clustering for future work.

\section[MG models and simulations]{Modified gravity models and $N$-body simulations}
\label{sec:simulations}

\subsection{Theoretical models}

In this subsection, we briefly describe the two modified gravity models analysed in this work, chameleon $f(R)$ gravity and the \ac{DGP} braneworld models \citep{DvaliDGP:2000PhLB..485..208D}. These are two of the most widely studied \ac{MG} models  and, as we  discuss below, are representative examples of two classes of screening mechanisms, which make them good test-beds for generic \ac{MG} models. For more detailed descriptions of these models, we refer the reader to \citet{Sotiriou:2010RvMP...82..451S,DeFelice:2010LRR....13....3D} for $f(R)$ gravity, and \citet{2003JCAP...11..014S,maartens2010brane} for \ac{DGP} models.

\subsubsection{$f(R)$ gravity}
\label{subsec:fRgrav}

The $f(R)$ gravity is a generalisation of Einstein's general relativity. In $f(R)$ gravity, the Einstein-Hilbert action in GR has an additional term, which is a function of the Ricci scalar $R$,
\begin{align}
    S = \int \dd^4 x \sqrt{-g} \qty{\frac{M^2_{\rm Pl}}{2} \qty[ R + f(R)] + \mathcal{L}_m } \ , \label{eqn:fRaction_sdv}
\end{align}
where $M_{\rm Pl} = (8 \pi G)^{-1/2}$ is the reduced Planck mass, $G$ is Newton's constant, $g$ is the determinant of the metric $g_{\mu \nu}$ and $\mathcal{L}_m$ the Lagrangian density for matter fields. 
Varying the action with respect to the metric $g_{\mu\nu}$ gives the modified Einstein equation,
\begin{align}
    G_{\mu \nu} + f_R R_{\mu \nu} - \qty(\frac{1}{2} f - \square f_R ) g_{\mu \nu} - \nabla_\mu \nabla_\nu f_R = 8\pi G T^m_{\mu \nu} \ , \label{eqn:einstein_field_eqn_for_fR_efsd}
\end{align}
in which 
\begin{equation}
G_{\mu\nu} \equiv R_{\mu\nu} - \frac{1}{2}g_{\mu\nu}R,
\end{equation} 
is the Einstein tensor, $f_R \equiv \dd f(R) / \dd R$, $\nabla_\mu$ is the covariant derivative corresponding to the metric $g_{\mu\nu}$, $\square \equiv \nabla^\alpha \nabla_\alpha$ and $T^m_{\mu\nu}$ is the energy momentum tensor for matter.

Eq.~\eqref{eqn:einstein_field_eqn_for_fR_efsd} is a fourth-order partial differential equation in $g_{\mu \nu}$. This equation can also be  considered as the standard Einstein equation in GR with a new dynamical degree of freedom, $f_R$, which is dubbed the  \textit{scalaron}  \citep[e.g.,][]{2011PhRvD..83d4007Z}. 
The equation of motion of $f_R$ can be obtained by taking the trace of Eq.~\eqref{eqn:einstein_field_eqn_for_fR_efsd}:
\begin{align}
    \square f_R = \frac{1}{3} \qty(R - f_R R + 2f + 8 \pi G \rho_m) \ , \label{eqn:fR_EoM_ewfasd}
\end{align}
where $\rho_m$ is the matter density.

For cosmological simulations in standard gravity, the Newtonian limit is commonly adopted. This includes the approximations that the  gravitational and scalar fields are weak (such that their higher-order terms can be neglected) and quasi-static (so that the time derivatives of the fields can be neglected compared to their spatial derivatives). Most modified gravity simulations (including the ones used in this work) adopt this assumption. In the context of $f(R)$ gravity and the Newtonian limit, the modified Einstein equation \eqref{eqn:einstein_field_eqn_for_fR_efsd} becomes
\begin{align}
    \boldsymbol{\nabla}^2 \Phi &\approx \frac{16 \pi G}{3} a^2 (\rho_m - \bar{\rho}_m) + \frac{1}{6} a^2 \qty[R(f_R) - \bar{R}] \ , \\
\intertext{and the equation of motion of the scalaron reduces to}
    \boldsymbol{\nabla}^2 f_R &\approx -\frac{1}{3} a^2 \qty[ R(f_R) - \bar{R} + 8 \pi G (\rho_m - \bar{\rho}_m)] \ , \label{eqn:fR_scalar_field_equation}
\end{align}
where $\Phi$ is the Newtonian potential, $\boldsymbol{\nabla}$ is the 3-dimensional gradient operator, and an overbar denotes the cosmic mean of a quantity.

In order to simulate cosmic structure formation in $f(R)$ gravity, one has to choose a specific functional form for $f(R)$. Here, we adopt the well-studied Hu-Sawicki model \citep{Hu:2007PhRvD..76f4004H}, but  generalise it slightly. The original functional form of $f(R)$ is
\begin{align}\label{eq:HS_fr}
    f(R) = -m^2 \frac{c_1 (-R/m^2)^n}{c_2 (-R/m^2)^n + 1} \ ,
\end{align}
where $m^2 \equiv \Omega_{m0} H_0^2$ and $c_1, c_2$ and $n$ are free model parameters. 
The parameter $n$ is a positive number, which is set to $n=1$ in most previous studies of this model \citep[however see e.g.,][for some examples of $n\neq1$]{Li:2011uw,Ramachandra:2020lue}. With this functional form, we have
\begin{align}\label{eq:generalised_HS_fr}
    f_R = -\left| \bar{f}_{R0} \right|\left(\frac{\bar{R}_0}{R}\right)^{n+1},
\end{align}
where $\bar{R}_0$, $\bar{f}_{R0}$ are, respectively, the present-day values of the background Ricci scalar and $\bar{f}_R$. Starting from this equation, we are able to consider also the case of $n=0$, which is not allowed by Eq.~\eqref{eq:HS_fr}. We will consider cases of $n=0,1,2$; for each $n$ we will consider a range of values of $f_{R0}$, to increase the diversity of model behaviour. For brevity, we will adopt the following nomenclature to label models: the model with $n=1$ and $-\log_{10}\left(|\bar{f}_{R0}|\right)=5$ will be called F5n1, and so on. 

The remaining free parameter of the theory is the background value of the scalar field $f_R$ at redshift $z = 0$, $\bar{f}_{R0}$. 
With a suitable choice of this parameter, $f(R)$ gravity recovers GR in high-density regions --- this is necessary to be consistent with solar system tests through the associated chameleon mechanism \citep{Khoury:2004PhRvL..93q1104K,Khoury:2004PhRvD..69d4026K}.
We show extensively the results of the model with $\bar{f}_{R0}=-10^{-5}$ and $n=1$, namely F5n1.
We note that a larger value of $|\bar{f}_{R0}|$ means a stronger deviation from standard gravity. The F5n1 model could be in slight tension with small-scale tests \citep[see, e.g.,][for a recent review of current cosmological\footnote{{Astrophysical constraints on this parameter are generally much stronger \citep[e.g.,][]{Desmond:2020gzn}, but they are in a different regime and have different systematic effects than cosmological constraints, and hence we shall not consider them here.}} constraints on $\bar{f}_{R0}$]{2014AnP...526..259L}. But since we aim to test gravity on much larger scales, it is nevertheless still a very valuable model to study: given its slightly stronger deviation from GR compared to models such as $|\bar{f}_{R0}| = 10^{−6}$ (F6n1), it can lead to important insights into how the deviations affect large-scale cosmological observables such as weak lensing and galaxy clustering statistics. 
In order to fully explore the GR testing capacities of upcoming large-scale structure survey, it is critical to gain a detailed understanding of how these measures are altered by possible modifications to gravity.

\subsubsection{Dvali-Gabadadze-Porrati (DGP) model}
\label{subsec:DGPmod}

In the braneworld model proposed by Dvali, Gabadadze and Porrati \citep{DvaliDGP:2000PhLB..485..208D}, the Universe is a four-dimensional brane embedded in a five-dimensional space-time (called the bulk). The gravitational action in this model is given by 
\begin{align}
    S = \int_{\text{brane}} \dd[4]{x} \sqrt{-g} \qty(\frac{R}{16 \pi G}) + \int_{\text{bulk}} \dd[5]{x} \sqrt{-g^{(5)}} \qty(\frac{R^{(5)}}{16 \pi G^{(5)}}) \ , \label{eqn:grav_act_DGP}
\end{align}
where a superscript $^{(5)}$ denotes the quantity in the five-dimensional bulk. This model has a self-accelerating branch of solution (sDGP), which gives a natural explanation for the cosmic acceleration (though with a distinctly different expansion history from $\Lambda$CDM), but the sDGP branch suffers from pathological problems \citep{2007CQGra..24R.231K} and its predictions have been found to be inconsistent with observations such as the \ac{CMB}, supernovae and local measurements of $H_0$ \citep[e.g.,][]{2007PhRvD..75f4003S,2008PhRvD..78j3509F}.

The so-called normal branch DGP (nDGP) gravity \citep{2007CQGra..24R.231K} cannot accelerate the Hubble expansion rate on its own and so to explain cosmological observations it is necessary to introduce an additional component of dark energy or a cosmological constant. This model is nevertheless still of interest as a useful toy model that features the Vainshtein screening mechanism  \citep{VAINSHTEIN1972393}. In this paper, we assume that there is an additional non-clustering dark energy component in this model, with which its expansion history is made identical to that of $\Lambda$CDM. The nDGP model provides an explanation why gravity is much weaker than the other fundamental forces \citep{maartens2010brane}: all matter species are assumed to be confined to the brane, while gravity could propagate through (leak into) the extra spatial dimensions. There is one new free parameter in the nDGP model, which can be defined as the ratio of $G^{(5)}$ and $G$, and it is known as the crossover scale, 
\begin{align}
    r_c \equiv \frac{1}{2} \frac{G^{(5)}}{G}\,.
\end{align}

Taking the variation of the DGP action, Eq.~\eqref{eqn:grav_act_DGP}, in a homogeneous and isotropic universe yields the modified Friedmann equation
\begin{align}
    \frac{H(a)}{H_0} = \sqrt{\Omega_{m0} a^{-3} + \Omega_{\rm DE}(a) + \Omega_{\rm rc}} - \sqrt{\Omega_{\rm rc}} \ , \label{eqn:Friedmann_eqn_nDGP}
\end{align}
where $\Omega_{\rm rc} \equiv 1 / (4H_0^2 r_c^2)$, and $\Omega_{\rm DE}$ is the density parameter of the additional dark energy component.
The dimensionless quantity $H_0 r_c$ can be used to quantify the departures from standard gravity.
If $H_0 r_c \to \infty$ then Eq.~\eqref{eqn:Friedmann_eqn_nDGP} returns to the $\Lambda$CDM case.
A larger value of $H_0 r_c$ means a weaker deviation from GR, because it means that the crossing scale $r_c$, above which gravity starts to have a non-standard 5-dimensional behaviour, is larger.

In the nDGP model, cosmological structure formation is governed by the modified Poisson and scalar field equations \citep{Koyama:2007ih},
\begin{equation}\label{eq:poisson_nDGP}
\nabla^2 \Phi = 4\pi G a^2 \delta \rho_{\rm m} + \frac{1}{2}\nabla^2\varphi\,,
\end{equation}
and
\begin{equation}\label{eq:phi_dgp}
\nabla^2 \varphi + \frac{r_c^2}{3\beta\,a^2c^2} \left[ (\nabla^2\varphi)^2
- (\nabla_i\nabla_j\varphi)^2 \right] = \frac{8\pi\,G\,a^2}{3\beta} \delta\rho_{\rm m}\,,
\end{equation}
where $\varphi$ is a new scalar degree of freedom, $\delta\rho_{\rm m} = \rho_{\rm m} - \bar{\rho}_{\rm m}$ and 
\begin{equation}\label{eq:beta_dgp}
\beta(a) \equiv 1 + 2 H\, r_c \left ( 1 + \frac{\dot H}{3 H^2} \right ) = 1 + \frac{\Omega_{\rm m}a^{-3} + 2\Omega_\Lambda}{2\sqrt{\Omega_{\rm rc}(\Omega_{\rm m}a^{-3} + \Omega_{\Lambda})}}\,.
\end{equation}

Here, we will study the nDGP model for various values of $H_0r_c$, and for easy references we will adopt the following rule: a model with $H_0r_c=1$ is called N1, and similarly for other values of $H_0r_c$.

\subsection{$N$-body simulations}
\label{subsec:Nbodydata}

In this subsection, we briefly introduce the $N$-body simulations we use to assess the performance of the \ac{RSD} models in the context of modified gravity theories.
Table~\ref{tab:simulation_summary} gives a summary of the simulation specifications. 
We focus on `main' or `distinct' dark matter haloes, and we leave a more detailed study for mock galaxies to a subsequent work.

\begin{table*}
\centering
\caption{The summary of the specifications of the simulations used in this work. Note that the \textsc{lightcone} simulations \citep[][]{2019MNRAS.483..790A} are only used in Appendix \ref{appendix:ConvergenceTests}, but we nevertheless include them here for completeness.}
\label{tab:simulation_summary}
    \begin{tabular}{cccccccccc}
    \toprule
     Simulation & Cosmology  & Code      & \makecell{Model \\  ($\times$ \#realisations)} & \makecell{Box size \\  ($h^{-1} \mathrm{Mpc}$) }  & \makecell{$M_{\text{particle}}$ \\ $(h^{-1} M_{\odot})$ } & $N_{\text{particle}}$ & \makecell{Force \\ resolution \\ ($h^{-1}{\rm kpc}$)} & \makecell{Halo \\ finder} & \makecell{Halo mass \\ definition} \\ \midrule
     \textsc{glam}  & {Planck15}   & {\textsc{glam}}    & {GR ($\times 72$)} & {$512.0$} & {$1.07 \times 10^{10}$} & {$1024^3$} & {$250.0$} & {\textsc{BDM}} & {$M_{\rm vir}$} \\\midrule
     \textsc{mg-glam}  & {Planck15}   & {\textsc{mg-glam}}    & \makecell{F5n0 ($\times 10$) \\ F5n1 ($\times 10$) \\ N1 ($\times 10$)} & {$512.0$} & {$1.07 \times 10^{10}$} & {$1024^3$} & $250.0$ & {\textsc{BDM}} & {$M_{\rm vir}$} \\\midrule
     \textsc{mg-glam}  & {Planck15}   & {\textsc{mg-glam}}    & \makecell{28 $f(R)$ models ($\times1$) \\ 29 nDGP models ($\times1$)} & {$512.0$} & {$1.07 \times 10^{10}$} & {$1024^3$} & $250.0$ & {\textsc{BDM}} & {$M_{\rm vir}$} \\\midrule
     \textsc{lightcone}          & Planck15   & \textsc{mg-gadget} & \makecell{GR ($\times 1$) \\ F5n1 ($\times 1$)} & $768.0$ & $4.50 \times 10^{9\phantom{0}}$ & $2048^3$ & $10.0$ & \textsc{subfind} & $M_{\rm 200c}$\\
     \bottomrule
    \end{tabular}
\end{table*}


In the main body of this paper, the simulations are run with \textsc{glam} \citep{2018MNRAS.478.4602K} (for $\Lambda$CDM) and its modified gravity extension, \textsc{mg-glam} \citep[][]{Hernandez-Aguayo:MG-GLAM2021,Ruan:2021MGGLAMfR} for all the \ac{MG} models. \textsc{glam} is a parallel particle-mesh (PM) code for the massive production of $N$-body simulations and mock galaxy catalogues in GR. It uses a regularly spaced 3D mesh of size $N_g^3$ covering the cubic simulation box of (comoving) volume $L_{\mathrm{box}}^3$. The force and mass resolution are defined by the size of a cell, $\Delta x = L_{\rm box} / N_g$, and the mass of each particle, 
\begin{align}
    M_{\rm particle} &= \Omega_{\rm m} \rho_{\rm c,0}  \frac{L_{\rm box}^3 }{N_{\rm particle}^3} \ ,
\end{align}
respectively, where $N_{\rm particle}^3$ is the number of simulation particles and $\rho_{\rm c,0} \equiv 3H_0^2 / (8 \pi G)$ is the present value of the critical density (see Appendix~A of \citealt{2018MNRAS.478.4602K} for details). \textsc{glam} solves the Poisson equation for the Newtonian potential in a cubic simulation box using the \ac{FFT} algorithm, and it uses the Cloud-In-Cell (CIC) scheme to implement the matter density assignment and force interpolation.

\textsc{mg-glam} extends \textsc{glam} to a general class of modified gravity theories by adding extra modules for solving MG scalar field equations. In the code papers of \textsc{mg-glam} \citep{Hernandez-Aguayo:MG-GLAM2021,Ruan:2021MGGLAMfR}, we describe the optimised multigrid relaxation algorithm used to solve the non-linear \ac{MG} equations, such as Eq.~\eqref{eqn:fR_scalar_field_equation} and Eq.~\eqref{eq:phi_dgp}, and their numerical implementations. In these papers we also reported some of the most interesting and basic cosmological quantities, such as the matter power spectrum and halo mass function, for several classes of \ac{MG} models, and compared these with the results from other high-precision \ac{MG} $N$-body codes, such as \textsc{ecosmog} \citep{2012JCAP...01..051L}, \textsc{mg-gadget} \citep[][]{2013MNRAS.436..348P} and the \ac{MG} modules of \textsc{arepo} \citep{2010MNRAS.401..791S,2019NatAs...3..945A,Hernandez-Aguayo:2020kgq}, finding good agreement.

In total, we have $72$ independent realisations of \textsc{glam} simulations for \ac{GR}, and we have simulated 30 $f(R)$ gravity models (with $10$ values of $\log_{10}|\bar{f}_{R0}|$ ranging between $-6$ and $-4.5$, respectively for $n=0,1,2$) and 30 nDGP models (for $30$ different values of $H_0r_c$, including N1) using \textsc{mg-glam}, with one realisation for each model. Moreover, for F5n0, F5n1 and N1, we have additional independent runs so that each of these models has $10$ realisations. All these runs adopt the $\Lambda$CDM cosmology with the Planck 2015 best-fitting cosmological parameters \citep[][hereafter Planck15]{Planck15Parameters:2016A&A...594A..13P}. The simulations follow the evolution of $1024^3$ dark matter particles in a simulation box with a side $L_{\rm box} = 512 \, h^{-1} \mathrm{Mpc}$, starting at an initial redshift of $z_{\rm init} = 100$ with the initial conditions (ICs) generated using the Zel'dovich approximation. For further details of these simulations see \cite{Hernandez-Aguayo:MG-GLAM2021,Ruan:2021MGGLAMfR}.

Table~\ref{tab:simulation_summary} presents the specifications of our simulations: the box size, particle mass $M_{\text{particle}}$, number of particles  $N_{\text{particle}}$, mesh numbers $N_g^3$, etc. The halo catalogues are produced using the bound density maxima (BDM) spherical overdensity halo finder \citep{2011ApJ...740..102K}. Only main haloes are studied in this work since the subhaloes are not well resolved due to the limited force resolution\footnote{However, in the code papers we found that the main haloes of these simulations are complete down to a halo mass of $\simeq 10^{12.5} \, h^{-1}M_\odot$. This should allow us to construct mock galaxy catalogues based on the halo occupation distribution (HOD) model \citep[see, e.g.,][]{Berlind:2002rn,Zheng:2004id}, though in this paper we will focus on haloes for simplicity, and leave a more detailed analysis using realistic mock galaxy catalogues to  future work.}. For the halo mass definition, the BDM halo finder adopts the virial mass $M_{\rm vir}$, which is the mass enclosed within a spherical overdensity of radius $R_{\rm vir}$, such that the mean overdensity within this sphere is $\Delta_{\rm vir} \approx 330$ times the mean matter density of the Universe. The virial overdensity is calculated according to \citet{1998ApJ...495...80B}. We saved halo catalogues at redshift $z = 0.0, 0.5$ and $1.0$ for analysis, and show the results of $z=0.5$ in the main text. The results obtained from the  other snapshots are presented in Appendix~\ref{appendix:more_cases}.

\begin{figure*}
    \centering
    \includegraphics[width=0.8\textwidth]{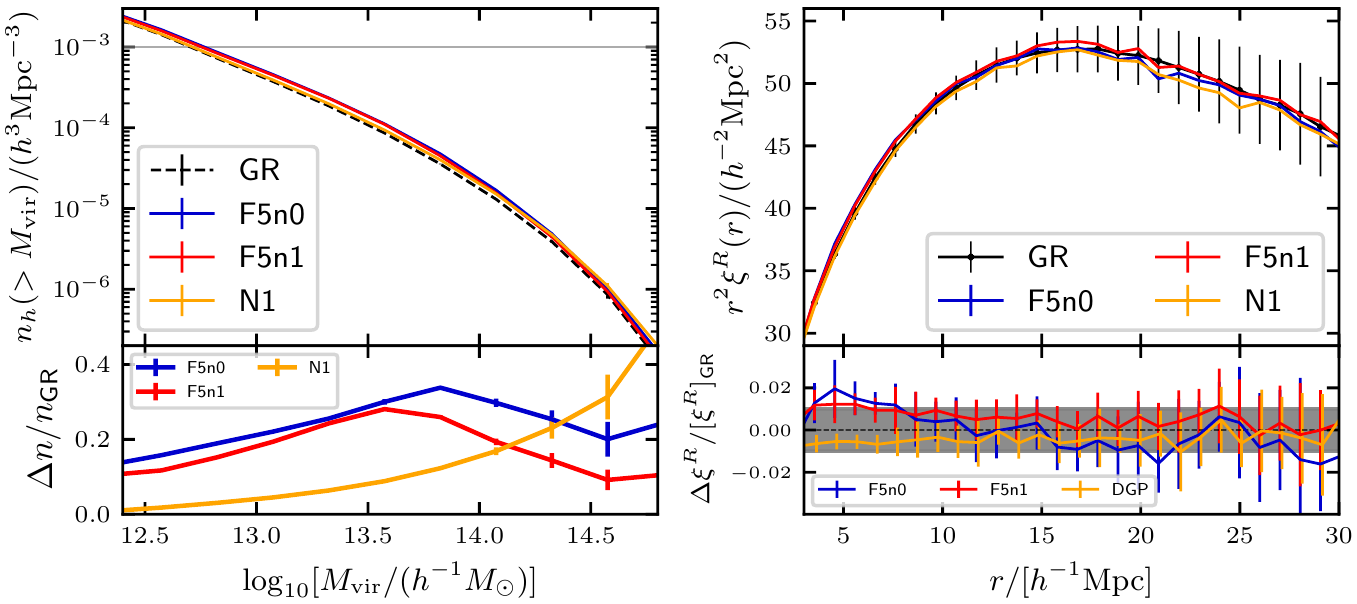}
    \caption{(Colour Online)
        Cumulative halo mass functions (left panel) and halo real-space correlation functions (right panel) at $z=0.5$, from (\textsc{mg-})\textsc{glam} simulations of the $f(R)$ model with $f_{R0}=-10^{-5}$, $n=0$ (F5n0, blue) and $n=1$ (F5n1, red), the \ac{DGP} model with $H_0 r_c = 1$ (N1, orange) and $\Lambda$CDM (black).
        The $\Lambda$CDM halo catalogues have fixed number density $n_h = 10^{-3} \, (h^{-1} \mathrm{Mpc})^{-3}$ (indicated by the grey horizontal line in the upper left panel) by selecting haloes more massive than a threshold value $M_{\rm min}$.
        For \ac{MG} halo catalogues, the mass cuts are tuned to match the $\Lambda$CDM halo correlation functions over a range of scales.
        The lower subpanels show the fractional difference between the \ac{MG} and $\Lambda$CDM results, with the grey shaded region in the lower right panel indicating $\pm1\%$. 
        The error bars present the standard deviation over $10$ realisations for each model ($72$ for $\Lambda$CDM results).
        Only the $\Lambda$CDM error bars are displayed in the upper left panel for clarity.
    }
    \label{fig:cHMF_DiffSim_comp}
\end{figure*}

In Fig.~\ref{fig:cHMF_DiffSim_comp}, we show some of the basic cosmological quantities predicted by the (\textsc{mg-})\textsc{glam} simulations. The left panel shows the cumulative \ac{HMF} for the GR, F5n0, F5n1 and N1 models, each showing the mean of 10 independent realisations; the lower subpanel shows the enhancements of the \ac{MG} models with respect to \ac{GR}, which agree very well with earlier simulation studies \citep[see][]{Hernandez-Aguayo:MG-GLAM2021,Ruan:2021MGGLAMfR}. The right panel of Fig.~\ref{fig:cHMF_DiffSim_comp} compares the real-space halo \ac{TPCF} of the same models; here we have tuned the number densities of the halo catalogues\footnote{Note that this tuning means that here we are not comparing halo catalogues in different models with exactly the same number density, but the latter is not our main interest anyway. On the other hand, as we shall see below, this tuning will make it easier when comparing other physical quantities.} such that the \ac{TPCF}s in all models agree with each other within $\simeq 1 \%$ between $5$ and $30 \, h^{-1}\mathrm{Mpc}$ (see the lower subpanel, which shows the relative difference from \ac{GR} after the tuning).

In Appendix \ref{appendix:ConvergenceTests}, we compare several halo pairwise velocity statistics predicted by these simulations with the predictions from a high-resolution simulation using a different code, and find that \textsc{mg-glam} gives reliable results down to small scales.

\section[the streaming model of RSD]{The streaming model of redshift space distortions}
\label{sec:STmodel}

In this section, we focus on the mapping between real-space and redshift-space two-point statistics. The redshift-space halo correlation function depends on the real-space correlation function and the halo pairwise velocity \ac{PDF}, as described by the so-called streaming model. We show that this \ac{PDF} can be accurately described by an \ac{ST} distribution in different \ac{MG} scenarios.
The \ac{ST} distribution has four free parameters, which can be related to the first four moments of the pairwise velocity \ac{PDF}. It can therefore match the skewness and kurtosis of the halo pairwise velocity \ac{PDF} predicted by $N$-body simulations. Finally, we will show that modelling skewness and kurtosis is relevant for constraining \ac{MG} models through small-scale redshift space clustering measurements. 

\subsection{Redshift-space distortions}
\label{subsect:rsd}

The real-space \ac{TPCF} is defined as 
\begin{equation}\label{eq:xi_r}
    \xi^{\rm R} (r) = \big\langle \delta (\bm{x}) \, \delta (\bm{x} + \bm{r}) \big\rangle \ ,
\end{equation} 
where $\delta(\bm{x})$ the number density contrast of the tracer field under investigation at position $\boldsymbol{x}$, and $\langle\cdots\rangle$ denotes the ensemble average. This quantity only depends on the length, $r$, of the pair separation vector, $\bm{r}$, due to the assumed statistical isotropy and homogeneity of clustering in real space.
$\xi^{\rm R} (r)$ describes the excess probability of finding a pair of tracers with separation $r$, compared with a random distribution of points.

In redshift space, the statistical isotropy is broken since there is a `special' line-of-sight direction, the velocity component of which could induce additional redshifts or blueshifts, causing distortions to the tracer correlation function Eq.~\eqref{eq:xi_r}.
This is known as redshift-space distortions, or RSD. 
In this space (denoted by superscript $^{\rm S}$), the correlation function, which is similarly defined as,
\begin{equation}\label{eq:xi_s}
    \xi^{\rm S} (s, \mu) = \big\langle \delta^{\rm S} (\bm{x}) \, \delta^{\rm S}(\bm{x} + \bm{s}) \big\rangle \ ,
\end{equation}
depends not only on the pair separation $s=|\boldsymbol{s}|$, but also on the angle of $\boldsymbol{s}$ with respect to the line-of-sight direction $\hat{\bm{z}}$, characterised by the cosine $\mu \equiv \hat{\bm{s}} \cdot \hat{\bm{z}}$.
This dependence can also be expressed by the separations perpendicular ($s_\perp$) and parallel ($s_{\parallel}$) to the line of sight, i.e., $\xi^{\rm S}(s_{\perp}, s_{\parallel})$, where $s=\sqrt{s^2_\parallel+s^2_\perp}$ and $s_\parallel=s\mu$, and we will use both notations.
It is convenient to decompose the 2D anisotropic correlation function $\xi^{\rm S}(s, \mu)$ into multipole moments in a basis of Legendre polynomials, $L_{\ell} (\mu)$, as 
\begin{align}
    \xi^{\rm S}(s, \mu) = \sum_{\ell} \xi_{\ell} (s) \, L_{\ell} (\mu) \ , \label{eqn:xiSsmu_decomp}
\end{align}
where $\ell$ is the order of the multipole. 
Odd $\ell$ moments vanish since $\xi^{\rm S} (s, \mu)$ is symmetric in $\mu$.
We will focus on the first three non-vanishing multipoles, i.e., the monopole ($\ell = 0$), the quadrupole ($\ell = 2$) and the hexadecapole ($\ell = 4$).
We measure $\xi^{\rm S}(s, \mu)$ in the separation range $4 \le s / (h^{-1}\mathrm{Mpc}) \le 30$ for $26$ linearly spaced bins of size $1\,h^{-1}\mathrm{Mpc}$, and in the angular cosine range $0 \le \mu \le 1$ for $240$ equally spaced linear bins. We have explicitly checked that these choices lead to converged result at subpercent level.

We use the publicly available Python package \textsc{Halotools}\footnote{\url{https://halotools.readthedocs.io/en/latest/}} \citep{2017AJ....154..190H} to measure real- and redshift space correlation functions of halo catalogues from the simulations. 
In cases where the simulation box is large enough, or where there are many independent realisations, we adopt the plane-parallel approximation, assuming that the line-of-sight direction $\hat{\bm{z}}$ is along one of the three axes of the Cartesian coordinate system for all haloes.
The systematic deviations caused by this assumption have been shown to be small for the current surveys \citep{2012MNRAS.420.2102S,2015MNRAS.447.1789Y}.
Under this approximation, the relation between the real ($\bm{r}$) and redshift ($\bm{s}$) space positions of a halo is given by 
\begin{align}\label{eq:z_space_coord}
    \bm{s} = \bm{r} + \frac{\bm{v} (\bm{r}) \cdot \hat{\bm{z}}}{a\, H(a)} \hat{\bm{z}} \ ,
\end{align}
where $\bm{v} (\bm{r})$ is the peculiar velocity of the halo and $H(a)$ is the Hubble factor at a given scale factor $a$.

\subsection{The streaming model}
\label{subsect:streaming}

The streaming model of \ac{RSD}, introduced by \citet{Peebles:1980lssu.book.....P} and subsequently generalised by \citet{1995ApJ...448..494F}, is a probabilistic approach to relate the clustering statistics of tracers in real and redshift spaces. 
The full complexities of modelling the redshift-space \ac{TPCF}, $\xi^{\rm S} (s_\perp, s_\parallel)$, are encoded in the pairwise velocity \ac{PDF}, which is the probability distribution of the relative velocities in a pair of tracers (haloes in our case) at a given halo separation; this is explicitly defined as $\bm{v}_{12} \equiv \bm{v}_2 - \bm{v}_1$, where $\bm{v}_{1},\bm{v}_{2}$ are the velocities of the two haloes in the pair. 


Because the number of tracers is conserved in real and redshift space, the fractional number overdensity in the two spaces is related by 
\begin{align}
    \big[1 + \delta^{\rm S} (\bm{s}) \big] \dd[3]{\bm{s}} = \big[1 + \delta^{\rm R} (\bm{r}) \big] \dd[3]{\bm{r}} \ .
\end{align}
This equation can be futher manipulated \citep{2004PhRvD..70h3007S} to obtain the exact relationship between real- and redshift-space two-point correlation functions,
\begin{align}
    1 + \xi^{\rm S} (s_\perp, s_\parallel) = \int_{-\infty}^{\infty} \dd r_\parallel \big[ 1 + \xi^{\rm R} (r) \big] \mathcal{P} (s_{\parallel} - r_{\parallel} | \bm{r}) \ , \label{eqn:streaming_model_core}
\end{align}
where 
\begin{equation}
    s_\perp \equiv r_\perp, \quad s_\parallel \equiv r_\parallel + \frac{v_\parallel}{aH(a)}, \label{eqn:spsp_rprp_relation_sdf}
\end{equation}
$\bm{r} \equiv (r_\perp, r_\parallel)$, $r \equiv \sqrt{r_\perp^2 + r_\parallel^2}$, and $\mathcal{P} (v_{\parallel} | \bm{r})$ is the \ac{PDF} of line-of-sight relative velocities of halo pairs separated by $\bm{r}$.

The line-of-sight pairwise velocity \ac{PDF}, $\mathcal{P} (v_{\parallel} | \bm{r})$ can be calculated from the full halo pairwise velocity distribution, $\mathcal{P} (v_r, v_t | r)$, where $v_r$ and $v_t$ are the pairwise velocity components parallel and transverse to the pair separation vector $\bm{r}$, respectively.
The line-of-sight projection of velocities is given by 
\begin{align}
    v_{\parallel} = v_r \cos \theta + v_t \sin \theta \ , \label{eqn:vpara_relation_vrt_rwfs}
\end{align} 
where $\theta \equiv \arctan(r_{\perp} / r_{\parallel})$ is the angle between the line of sight and the separation vector $\bm{r}$.
Therefore,
\begin{align}
    \mathcal{P} (v_{\parallel} | \bm{r}) = \int \frac{\dd{v_r}}{\sin \theta} \mathcal{P} \qty(\left.v_r, v_t=\frac{v_{\parallel} - v_r \cos \theta}{\sin \theta} \right| r) \ .
\end{align}
Note that the distribution $\mathcal{P} (v_r, v_t | r)$ only depends on the separation length $r$ (instead of the vector $\bm{r}$) due to statistical homogeneity and isotropy in real space.
It is an intrinsic property of $N$-body systems, which are determined by dynamical evolution under gravity.

Since we are discussing halo velocity fields, rather than the velocities of randomly chosen points in space, the moments $m_{ij}$ and the central moments $c_{ij}$ (where $i,j$ are non-negative integers) of the pairwise velocity \ac{PDF}, which are defined as 
\begin{align}
    m_{ij}(r) &\equiv \int \dd{v_r} \dd{v_t} (v_r)^i (v_t)^j \mathcal{P} (v_r, v_t | r) \ , \label{eqn:mij_def0}
\end{align}
and 
\begin{align}
    c_{ij}(r) &\equiv \int \dd{v_r} \dd{v_t} \qty[v_r - m_{10}(r)]^i \qty[v_t - m_{01}]^j \mathcal{P} (v_r, v_t | r) , \label{eqn:cij_def0}
\end{align}
should be weighted by halo mass when measured from simulations,
\begin{align}\label{eq:mij_def}
    m_{ij} (r) &= \frac{\left\langle [1 + \delta(\bm{x}_1)][1 + \delta (\bm{x}_2)] (v_r)^i (v_t)^j \right\rangle}{\left\langle [1 + \delta(\bm{x}_1)] [1 + \delta (\bm{x}_2)] \right\rangle} \ ,
\end{align}
and 
\begin{align}
    c_{ij} (r) &= \frac{\left\langle    [1 + \delta(\bm{x}_1)][1 + \delta (\bm{x}_2)] \qty[v_r - m_{10}(r)]^i \qty[v_t - m_{01}(r)]^j  \right\rangle}{\left\langle [1 + \delta(\bm{x}_1)] [1 + \delta (\bm{x}_2)] \right\rangle}, \label{eqn:cij_def}
\end{align}
where $r \equiv |\bm{x}_2 - \bm{x}_1|$.
Statistical isotropy in the transverse plane implies that only moments with even powers of the transverse component are non-zero. 
The four lowest order non-zero moments are 
\begin{align}
\begin{split}
    \hspace{2cm}& m_{10}, \\
    & c_{20}, c_{02}, \\ 
    & c_{30}, c_{12}, \\ 
    & c_{40}, c_{22}, c_{04} \ .
\end{split}
\end{align}

Similarly, the line-of-sight velocity moments and central moments are defined as 
\begin{align}
    m_n (\bm{r}) &{\color{red}\equiv} \int \dd{v_{\parallel}} (v_{\parallel})^n \,  \mathcal{P} (v_{\parallel} | \bm{r})  \\
\intertext{and}
    c_n (\bm{r}) &{\color{red}\equiv} \int \dd{v_{\parallel}} \qty[v_{\parallel} - m_1 (\bm{r})]^n \, \mathcal{P} (v_{\parallel} | \bm{r}) \ .
\end{align}
According to Eq.~\eqref{eqn:vpara_relation_vrt_rwfs}, the relations between the moments of the PDFs $\mathcal{P}\left(v_\parallel|\boldsymbol{r}\right)$ and $\mathcal{P}\left(v_r,v_t|r\right)$ are given by \citep{Cuesta-Lazaro:2020MNRAS.498.1175C}
\begin{equation}\label{eq:moments_relation}
c_n (r_\perp, r_\parallel) = \sum_{k=0}^n \binom{n}{k} \mu^k (1 - \mu^2)^{\frac{1}{2}(n-k)} c_{k, n-k}(r), 
\end{equation}
where $\mu \equiv r_{\parallel}/r$ is the direction cosine as before, $c_n$ is the $n$-th central moment of the line-of-sight velocity distribution $\mathcal{P}(v_\parallel | r_\perp, r_\parallel)$, and $c_{k, n-k}$ denotes the moment of the $k$-th radial component, $(n-k)$-th transverse component of $\mathcal{P}(v_r,v_t|r)$. The $n$-th moment about the origin is denoted as $m_n$. In what follows, we will need the line-of-sight pairwise velocity moments $m_1,c_{2\text{-}4}$ for the streaming model predictions, and these will be obtained by first measuring the pairwise velocity moments $c_{ij} (r)$ from the simulated halo catalogues in real space and then performing the conversions using Eq.~\eqref{eq:moments_relation}, since the latter are an intrinsic property of halo catalogues while the former also depend on the specified line-of-sight direction.


\subsubsection{The Gaussian streaming model (GSM)}

In its early applications, the streaming model was used to predict the galaxy clustering measured from the CfA survey \citep{1983ApJ...267..465D}. In this case the best fit to the observational data was found using an exponential form for the  pairwise velocity distribution.
\citet{1995ApJ...448..494F} showed that the streaming model with a Gaussian velocity \ac{PDF} and a scale-dependent velocity dispersion could reproduce the linear perturbation theory result for \ac{RSD} on large scales. \citet{2004PhRvD..70h3007S} demonstrated that the pairwise velocity \ac{PDF} is not Gaussian, even for a Gaussian matter density field, but it can be approximated by a Gaussian near its peak.
Based on a non-perturbative resummation of the linearised limit \citep{1995ApJ...448..494F} of the streaming model equation~\eqref{eqn:streaming_model_core}, \citet{Reid:2011MNRAS.417.1913R,Carlson:2013MNRAS.429.1674C} proposed that the line-of-sight pairwise velocity \ac{PDF} can be approximated by a Gaussian function,
\begin{align}\label{eq:gaussian_pdf}
    \mathcal{P}_{\rm G} (v_{\parallel} | \bm{r}) = \frac{1}{\sqrt{2\pi c_2 (\bm{r})}} \exp \qty[- \frac{\qty(v_{\parallel} - m_1 (\bm{r}))^2}{2 c_2 (\bm{r})}] \,,
\end{align} 
where we note that the Gaussian model parameters $m_1$ and $c_2$ are scale-dependent. The \ac{GSM} has become one of the most commonly used \ac{RSD} models in galaxy surveys \citep[e.g.][]{2012MNRAS.426.2719R,2014MNRAS.439.3504S,2017MNRAS.469.1369S,2020MNRAS.499.5527T}.

Considering the massive dark matter haloes, \citet{2018MNRAS.479.2256K} applied the \ac{CLPT} formalism to compute the ingredients in the \ac{GSM}, including the real-space clustering, and the first and second order pairwise velocity moments. They obtained predictions for the redshift-space correlation function monopole and quadrupole which are accurate to $2 \text{-} 4$ per cent down to $\simeq25 \, h^{-1}\mathrm{Mpc}$, compared to statistics measured in $N$-body simulations. 
\citet{Bose:2017JCAP...08..029B} adopted the \ac{GSM} combined with the regularised perturbation theory to compute the large-scale redshift-space halo power spectrum and \ac{TPCF},  for Vainshtein screened and Chameleon screened \ac{MG} models as well as \ac{GR}. 
Also based on the \ac{GSM}, \citet{Bose:2020JCAP...09..001B} presented a hybrid approach to predict the quasi non-linear redshift space matter power spectrum multipoles. 
\citet{2020JCAP...01..055V} extended the \ac{GSM} to calculate the redshift-space correlation functions for biased tracers in modified gravity models, by employing the \ac{LPT} and \ac{CLPT} resummation scheme to predict the ingredients of the \ac{GSM}, including $\xi^R(r)$, $m_{1}(\bm{r})$ and $c_2 (\bm{r})$.
Their new approach qualitatively reproduces the redshift-space correlation function quadrupole in the \ac{MG} simulations compared, down to at least $17 \, h^{-1}\mathrm{Mpc}$, and traces the shape of the hexadecapole down to similar small scales.

\begin{figure*}
    \centering
    \includegraphics[width=0.96\textwidth]{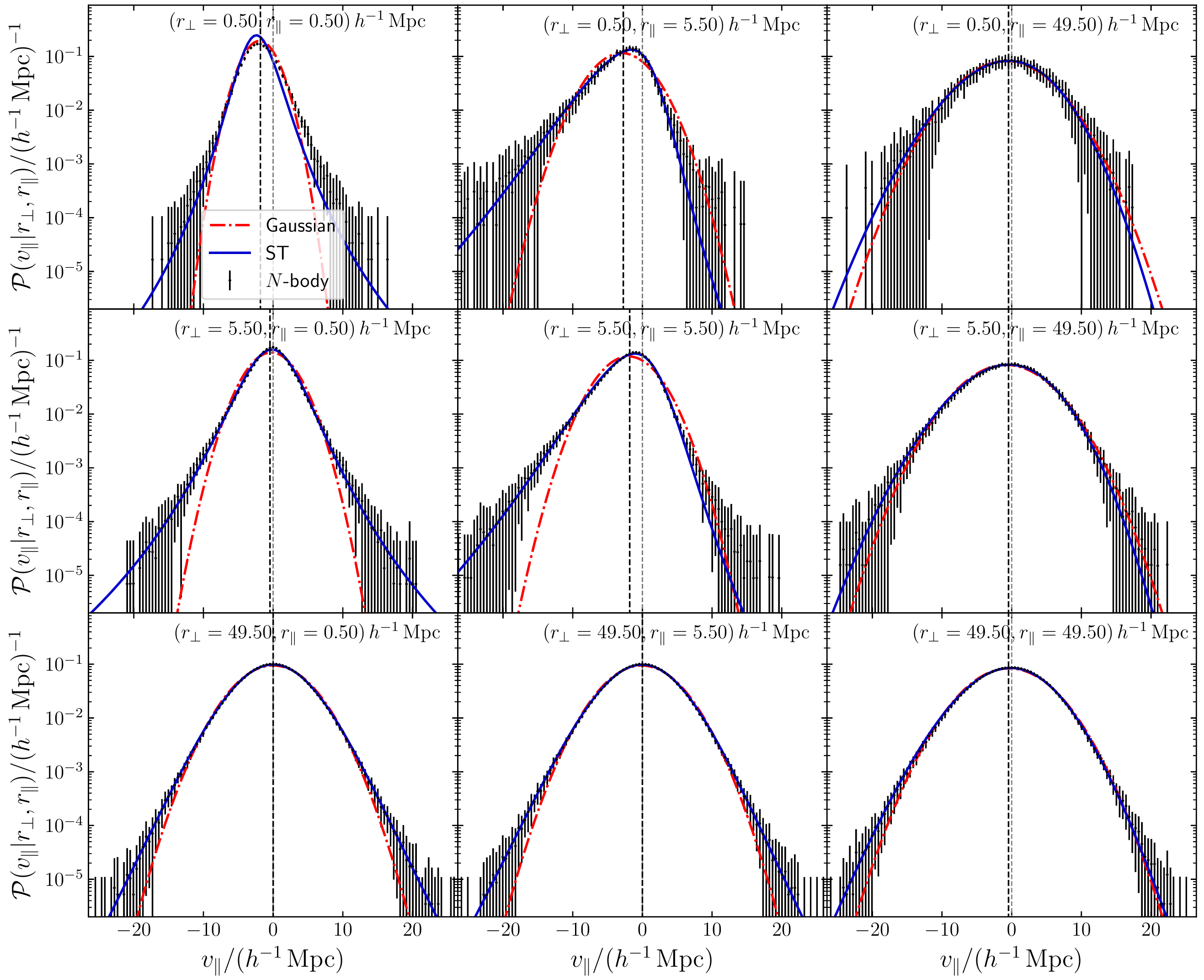}
    \caption{(Colour Online) The pairwise line-of-sight velocity distribution $\mathcal{P} (v_{\parallel} | \bm{r})$ for dark matter haloes with  number density $\bar{n}_h = 10^{-3} (h^{-1} \mathrm{Mpc})^{-3}$ at $z=0.5$, evaluated at different separations $\bm{r} = (r_{\perp}, r_{\parallel})$, from ten \textsc{mg-glam} cosmological runs for the F5n1 model.
    The rows show increasing $r_{\parallel}$ separation, and the columns show increasing $r_{\perp}$.
    The black dots with error bars represent the mean and standard deviation from ten realisations. 
    The red dash-dotted and blue solid lines show the Gaussian and ST models, respectively.
    The best-fitting parameters are obtained by converting the measured pairwise velocity moments, instead of directly fitting the black dots, as described in Section~\ref{subsect:streaming}.
    The black dashed lines represent the line-of-sight pairwise velocity mean, $m_{1} (\bm{r})$, by integrating the measured velocity PDF, and the grey dashed lines show zero velocity value to aid visualisation. 
    All velocities are rescaled by $1/(aH)$ according to Eq.~\eqref{eq:z_space_coord} so that they have the unit of length.}
    \label{fig:VlosPDF_MGGLAM_fRn1_NumDen3.0}
\end{figure*}

Despite its simplicity and popularity, it is well known that the Gaussian model described by Eq.~\eqref{eq:gaussian_pdf} does not fully describe the pairwise velocity \ac{PDF}, especially for pairs at small separations (\citet{2015MNRAS.446...75B,2015PhRvD..92f3004U,2016MNRAS.463.3783B,2018MNRAS.479.2256K}, see also Fig.~\ref{fig:VlosPDF_MGGLAM_fRn1_NumDen3.0}),because the true \ac{PDF} as measured from simulations can have significant skewness and kurtosis, which are absent in a Gaussian \ac{PDF}.

\subsubsection{The Skewed Student-t (ST) distribution}

\citet{Cuesta-Lazaro:2020MNRAS.498.1175C} proposed to use the so-called \ac{ST} distribution \citep{2009arXiv0911.2342A} to model $\mathcal{P}\left(v_\parallel|\bm{r}\right)$ \citep[see also][for an earlier application in a similar context]{10.1093/mnras/stt411}.
The \ac{ST} distribution is constructed from the Student's $t$-distribution, whose \ac{PDF} for a random variable $x$ in one dimension is given by 
\begin{align}
    t_1 (x-x_c | w, \nu) = \frac{\Gamma (\frac{\nu+1}{2})}{\sqrt{\nu \pi} w \Gamma(\frac{\nu}{2}) } \qty[1 + \frac{1}{\nu} \qty(\frac{x - x_c}{w})^2]^{-\frac{\nu+1}{2}}  . \label{eqn:tdist_pdf_1d_sfv}
\end{align}
This distribution is characterised by three parameters: the  location of the centre $x_c$, the shape parameter $w$, and the number of degrees of freedom $\nu$.

The expression for the ST distribution of line-of-sight pairwise velocities, which originates from the $t$-distribution \eqref{eqn:tdist_pdf_1d_sfv}, is given by 
\begin{align}
    &\mathcal{P}_{\rm ST} \qty(v_{\parallel}; {\color{gray}  v_c (\bm{r}), w (\bm{r}), \alpha (\bm{r}), \nu (\bm{r})} | \bm{r}) = \frac{2}{w} t_1 (v_{\parallel} - v_c | 1, \nu) \notag \\
    &  \times T_1\Bigg(
        \alpha \frac{v_{\parallel} - v_c}{w} \bigg[ \frac{\nu + 1}{\nu + \left( ({v_{\parallel} - v_c})/{w}\right)^2} \bigg]^{1/2}; \nu + 1 \Bigg) \ , \label{eq:st_pdf}
\end{align}
where $T_1$ is the one-dimensional cumulative $t$-distribution with $\nu+1$ degrees of freedom, and $v_c,w,\alpha$ and $\nu$ are the four free parameters, themselves functions of $\bm{r}$, which fully specify the \ac{ST} distribution. 

Although Eq.~\eqref{eq:st_pdf} looks quite lengthy, it has the advantage that the four parameters can be analytically related to its first four moments $m_1$ and $c_{2,3,4}$ \citep[see Eqs.~(A1)-(A6) in Appendix A of][we have reproduced these relations in Appendix \ref{appendix:app1} of this paper for completeness]{Cuesta-Lazaro:2020MNRAS.498.1175C}. Furthermore,  \cite{Cuesta-Lazaro:2020MNRAS.498.1175C} found that the \ac{ST} distribution fits the  $\mathcal{P}\left(v_\parallel|\bm{r}\right)$ measured from simulations very well, in particular for close pairs, and consequently it leads to much more accurate predictions of the RSD multipoles $\xi^{\rm S}_\ell(s)$ at small scales. We will see that it also works very well for the modified gravity models described in Sections~\ref{subsec:fRgrav} and \ref{subsec:DGPmod}.

In practice, the \ac{ST} model parameters, $v_c (\bm{r}), w (\bm{r}), \alpha (\bm{r}), \nu (\bm{r})$, are determined in the following way: 
(1) measure the lowest four moments of the pairwise velocity distribution $\mathcal{P}\left(v_r, v_t | r\right)$ from the simulated halo catalogues, (2) convert these to the lowest four moments of the line-of-sight projected velocity \ac{PDF} $\mathcal{P}\left(v_\parallel|\bm{r}\right)$ using Eq.~\eqref{eq:moments_relation}, and (3) compute $v_c,w,\alpha,\nu$ using the relations given in Appendix \ref{appendix:app1}. Schematically this can be illustrated as follows:
\begin{align}
        &\qty{m_{10}; c_{20, 02}; c_{30,12}; c_{40,22,04}} (r) \notag \\
        \xrightarrow{\text{Eq.~\eqref{eq:moments_relation}}} &\qty{m_1, c_2, c_3, c_4}(\bm{r}) \notag \\
        \xrightarrow{\text{Appendix~\ref{appendix:app1}}} &\qty{v_c, w, \alpha, \nu}(\bm{r}) \ . \label{eqn:method_ST_best_params_fitting}
\end{align}

\section{Results}
\label{sec:res}

Our aim is to demonstrate the importance of modelling higher order moments, beyond the mean and variance of the pairwise velocity distribution, to constrain modified gravity models using \ac{RSD}. In this section, we show that using the \ac{ST} distribution as a generic phenomenological model to convert the velocity moments into redshift-space clustering through the streaming model is also applicable to \ac{MG} models, by showing that its prediction agree very well with the measurements from $N$-body simulations, down to scales around $5 \, h^{-1}\mathrm{Mpc}$. 

In Sect.~\ref{subsec:sming_res} we analyse the ingredients of the \ac{SM}s, in particular the halo pairwise line-of-sight velocity PDF and its moments, as measured from (\textsc{mg-})\textsc{glam} simulations for a range of \ac{MG} models. In Sect.~\ref{subsec:halo_2pcf_res} we show that the \ac{STSM} can accurately predict both the redshift-space \ac{TPCF} multipoles in different \ac{MG} models, but also their relative enhancement with respect to \ac{GR}, down to small scales. In Sect.~\ref{subsec:fisher} we perform a simple Fisher analysis to illustrate how, by including small-scale RSD information, the power of galaxy clustering analyses in constraining \ac{MG} models can be significantly improved.

\subsection{Streaming model ingredients}
\label{subsec:sming_res}

\subsubsection{Halo line-of-sight pairwise velocity PDFs}

In Fig.~\ref{fig:VlosPDF_MGGLAM_fRn1_NumDen3.0}, we show the line-of-sight pairwise velocity PDF of dark matter haloes from the F5n1 simulations run using \textsc{mg-glam}, for nine selected combinations of $(r_{\perp}, r_{\parallel})$ covering large, intermediate and small scales. The figure shows increasing $r_{\perp}$ values from top to bottom and increasing $r_{\parallel}$ values from left to right. The black dots represent the measured \ac{PDF}s of dark matter haloes, and the lines show the Gaussian (red) and ST (blue) distributions. The best-fitting model parameters are obtained by converting the measured pairwise velocity moments, as described in Section~\ref{subsect:streaming} and specifically in Eq.~\eqref{eqn:method_ST_best_params_fitting} for the ST model. Comparing this plot with Fig.~2 of \citet{Cuesta-Lazaro:2020MNRAS.498.1175C}, which shows measurements from the standard gravity simulation suite \text{Dark Quest} \citep{2019ApJ...884...29N}, we see that the same conclusions can be reached regarding the relative performance of the Gaussian and ST models, even though  several aspects  of the analyses are different, such as: the gravity models ($f(R)$ gravity versus $\Lambda$CDM), the $N$-body codes used, the halo finders applied and the mass definitions.

\begin{figure*}
    \centering
    \includegraphics[width=0.9\textwidth]{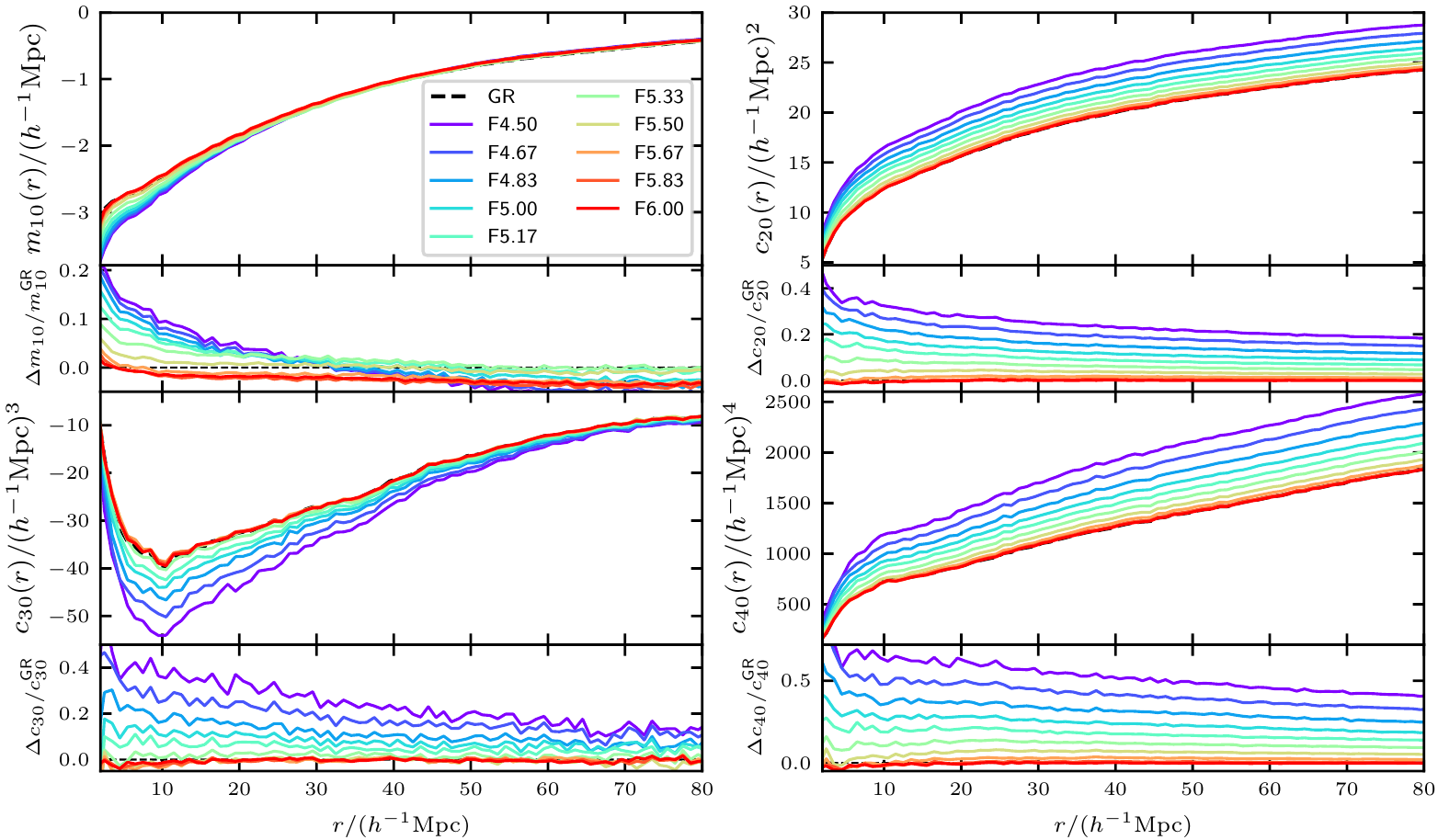}
    \caption{(Colour Online) The four lowest-order moments of the radial and transverse halo pairwise velocity distribution for the $f(R)$ gravity model with $n = 1$ and $10$ values of $\log_{10} |f_{R0}|$ equally spaced in the range between $-6.0$ and $-4.5$, as labelled, at $z = 0.5$, from the \textsc{mg-glam} simulations. The lower subpanels show the relative differences between the $f(R)$ and \ac{GR} models. The horizontal dashed line denotes $0$. The halo catalogues have a fixed number density of $n_h=10^{-3}~\left(h^{-1}\mathrm{Mpc}\right)^{-3}$ for all models.
    }
    \label{fig:moments_MGGLAM_fR_erwsfdcv2}
\end{figure*}

\begin{figure*}
    \centering
    \includegraphics[width=0.9\textwidth]{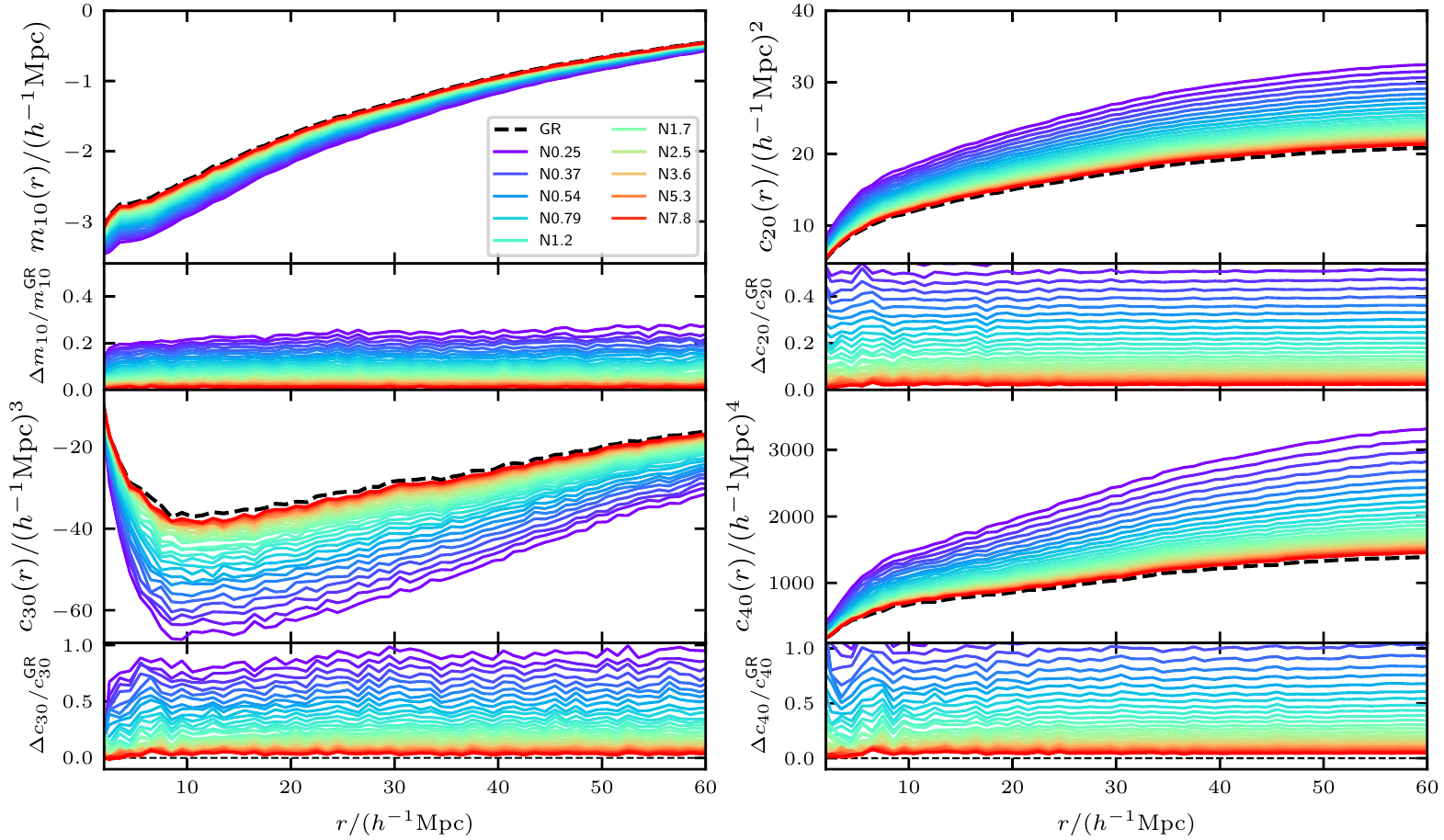}
    \caption{(Colour Online) The same as Fig.\ref{fig:moments_MGGLAM_fR_erwsfdcv2}, but for the \ac{DGP} model with $30$ $H_0 r_c$ values in the range of $[0.25, 10]$, as labelled.}
    \label{fig:moments_MGGLAM_DGP_erwsfdcv2}
\end{figure*}

\begin{figure*}
    \centering
    \includegraphics[width=0.9\textwidth]{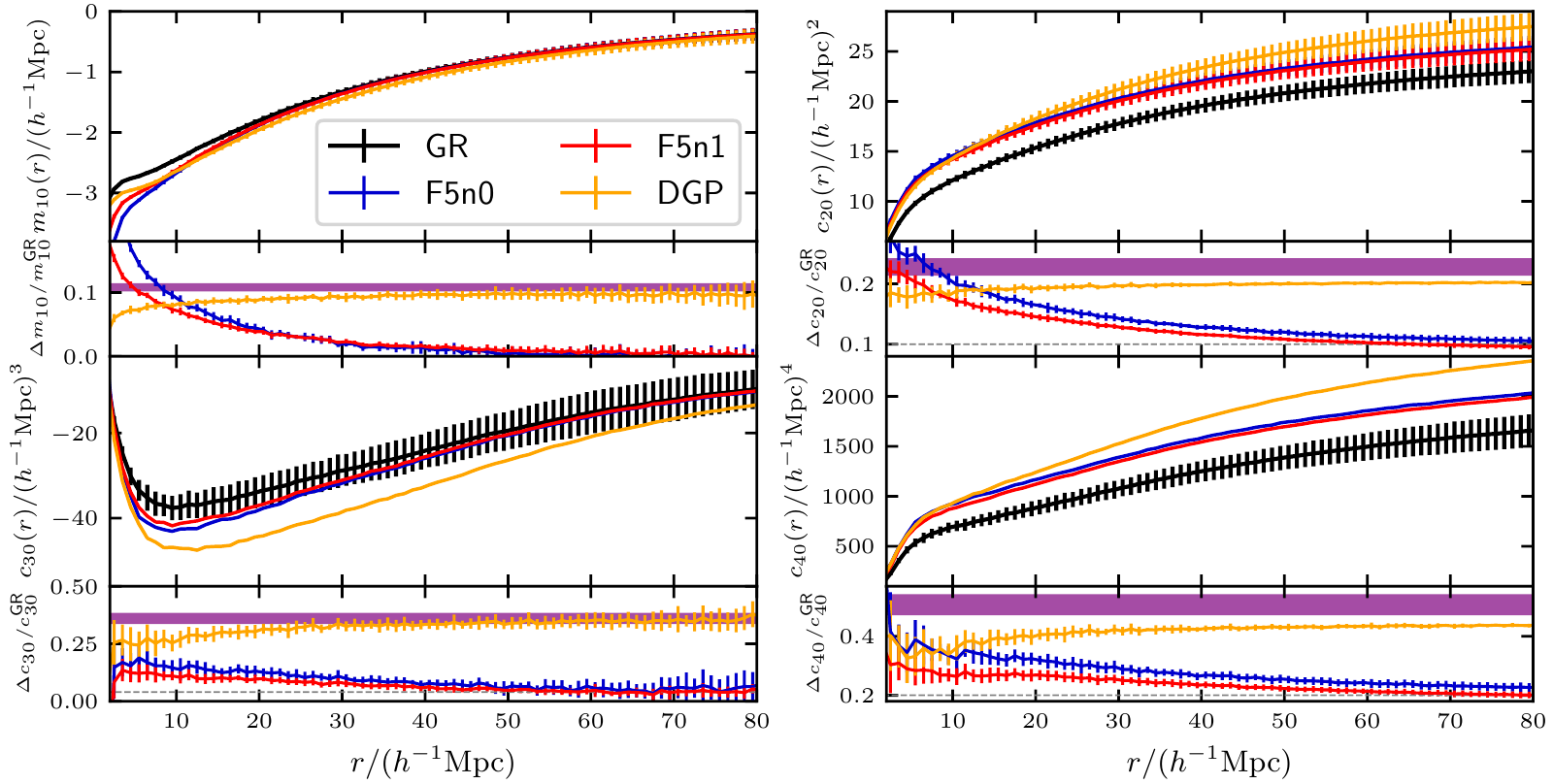}
    \caption{(Colour Online) The four lowest order moments of the radial and transverse halo pairwise velocity distribution at $z = 0.5$, for the $f(R)$ model with $f_{R0} = -10^{-5}$, $n = 0$ (F5n0, blue) and $1$ (F5n1, red), the \ac{DGP} model with $H_0 r_c = 1$ (N1, orange) and the $\Lambda$CDM (black) model. The lower subpanels show the relative differences     between the \ac{MG} and \ac{GR} models. The error bars present the standard deviation of $10$ realisations for each model. For the third and fourth order moments we only show the error bars of the $\Lambda$CDM results to avoid clutter. The purple bands show the theoretical prediction by Eq.~\eqref{eq:c_enhancement_lin}. This figure differs from Figs.~\ref{fig:moments_MGGLAM_fR_erwsfdcv2} and \ref{fig:moments_MGGLAM_DGP_erwsfdcv2} in that here the number densities of haloes have been tuned slightly so that the \ac{MG} models all match the real-space halo \ac{TPCF} of \ac{GR} in the range $r\in[5,30] \, h^{-1}\mathrm{Mpc}$.
    }
    \label{fig:moments_AllModelsWith10Res_NumDen3p0}
\end{figure*}

In Section~3.2 of \citet{Cuesta-Lazaro:2020MNRAS.498.1175C} there is a comprehensive discussion on how the behaviour of $P(v_{\parallel} | \bm{r})$ depends on pair separation. Here we make similar observations:
\begin{itemize}
    \item The ST model is a better description of the simulation measurements than the Gaussian distribution on all scales we have looked at, at the expense of requiring two more parameters to quantify the skewness and kurtosis of the \ac{PDF}. The improvement is significant for small separations ($r_{\perp}, r_{\parallel} \lesssim 25 \, h^{-1}\mathrm{Mpc}$).
    \item The behaviour of $P(v_{\parallel} | \bm{r})$ in \ac{GR} and $f(R)$ gravity is qualitatively similar, although the gravity is enhanced in the latter. The \ac{ST} model achieves a similar level of agreement as in the \ac{GR} case. Since the \ac{ST} distribution describes the measured velocity PDF accurately, we can use the ``best-fit''\footnote{Note that quotation marks are used here since strictly speaking this is not a fit. Instead, as described above, the \ac{ST} distribution parameters have been calculated directly using the measured velocity moments.} \ac{ST} results as proxies to explore the differences in $\mathcal{P} (v_{\parallel} | \bm{r})$ between $f(R)$ gravity and $\Lambda$CDM.
    \item For small separations, such as $r_{\parallel} = 5.50 \, h^{-1} \mathrm{Mpc}$, $r_{\perp} = 0.5$ or $5.50 \, h^{-1} \mathrm{Mpc}$, the line-of-sight velocity distributions are strongly skewed towards negative pairwise velocities.
    This can be explained by the fact that such close halo pairs are more likely to be located in high-density regions where haloes infall towards each other ($v_{\parallel} < 0$).
    This skewness is less obvious when we go to large $r_{\perp}$ \textit{or} $r_{\parallel}$ (e.g., $49.50 \, h^{-1} \mathrm{Mpc}$), since for large separations the probabilities of finding infalling and receding halo pairs tend to differ less. 
    \item The measured line-of-sight velocity \ac{PDF}s are heavily tailed compared with their best-fit Gaussian ones.
\end{itemize}

\subsubsection{Halo pairwise velocity moments}
\label{sec:halo_vel_moments}

Fig.~\ref{fig:moments_MGGLAM_fR_erwsfdcv2} shows the four lowest-order moments of the radial and transverse halo pairwise velocity \ac{PDF}s from halo catalogues at $z = 0.5$ with fixed number density $\bar{n}_h = 10^{-3} (h^{-1}\mathrm{Mpc})^{-3}$ for the $f(R)$ gravity model with $n=1$ and ten $\log_{10} |f_{R0}|$ values evenly spaced between $-6.0$ (the weakest modification) and $-4.5$ (the strongest), along with the relative differences with respect to the $\Lambda$CDM results displayed in the lower subpanels. We have checked the results for $f(R)$ models with $n=0$ and $n=2$, and found similar results, but for clarity those are not shown here. Fig.~\ref{fig:moments_MGGLAM_DGP_erwsfdcv2} is the same as Fig.~\ref{fig:moments_MGGLAM_fR_erwsfdcv2}, but presents the \ac{DGP} model with $30$ $H_0 r_c$ values logarithmically spaced between $0.25$ and $10$. We only show one of each higher-order moment to make the plot easier to read.

The differences in the velocity moments between the \ac{MG} and \ac{GR} models are caused by the \ac{MG} effects and the different halo populations. The latter occurs because, at fixed halo number density, the haloes from different models are likely not to have a one-to-one correspondence, even though the simulations start from the same ICs. For example, the contribution of the $f(R)$ gravity effect is suppressed on scales larger than the range of the fifth force. Therefore, we expect that the radial mean velocity relative enhancement, $\Delta m_{10} / m_{10}^{\rm GR}$, tends to be zero on large scales (e.g., $\gtrsim 40 \, h^{-1}\mathrm{Mpc}$). However, due to the halo population difference, we see that this is not the case as shown in the upper left panel of Fig.~\ref{fig:moments_MGGLAM_fR_erwsfdcv2}: $\Delta m_{10} / m_{10}^{\rm GR}$ on large scales is scale-independent but $\bar{f}_{R0}$-dependent. We are mainly interested in the difference caused by \ac{MG} effects; however, Figs.~\ref{fig:moments_MGGLAM_fR_erwsfdcv2} and \ref{fig:moments_MGGLAM_DGP_erwsfdcv2} only provide an incomplete picture of how the velocity moments depend on the \ac{MG} model parameters.

We can isolate the \ac{MG} effects, at least on large scales, on halo pairwise velocity moments and halo clustering by tuning the mass cut of \ac{MG} halo catalogues, so that the real-space correlation functions $\xi^R (r)$ agree with the \ac{GR} ones on large scales.
Due to the small simulation box size, the correlation functions are noisy and the tuning is not reliable for a single realisation.
We only tuned the \ac{MG} models for which we have run ten or more realisations, i.e. \ac{GR}, F5n0, F5n1 and N1.
The resulting real-space correlation functions are presented in the right panel of Fig.~\ref{fig:cHMF_DiffSim_comp}.
In the rest of this paper, we will always use these tuned \ac{MG} halo catalogues for the halo clustering analysis unless otherwise stated.

Fig.~\ref{fig:moments_AllModelsWith10Res_NumDen3p0} shows the same halo velocity moments measurements as in Figs.~\ref{fig:moments_MGGLAM_fR_erwsfdcv2} and \ref{fig:moments_MGGLAM_DGP_erwsfdcv2}, but for the matched halo catalogues and models only. We can see that after matching the large scale halo correlation functions, the mean velocity relative difference of $f(R)$ gravity with respect to $\Lambda$CDM is consistent with zero on scales $\gtrsim 40 \, h^{-1}\mathrm{Mpc}$, for both F5n1 and F5n0. The different values of $n$ only affect the small-scale mean velocities, with the boost in $n=0$ being larger as the fifth force is less screened in this case \citep[cf. Sect.~5.1 of ][]{Ruan:2021MGGLAMfR}. For higher order moments, the relative differences on large scales converge toward non-zero constants, whose values are slightly different for $n = 0$ and $1$. For the nDGP model (N1), the behaviour of the velocity moments is qualitatively different from that of $f(R)$ gravity on all scales. We will interpret this result in the context of \ac{MG} effects next.

As mentioned in Section \ref{sec:simulations}, the \ac{DGP} and $f(R)$ gravity models feature different screening mechanisms (Vainshtein vs.~thin-shell chameleon screening). 
In the DGP case, the fifth force is screened close to and inside massive bodies, but is unscreened and proportional to the Newtonian force when placed at a far distance. This means that structure formation is enhanced on large scales here, which is in contrast to $f(R)$ gravity, where the finite range of the fifth force means that structure formation is enhanced only below the Compton wavelength of the scalaron field. 
As a result, unlike in $f(R)$ gravity, the large-scale value of the radial mean velocity enhancement in \ac{DGP} is non-zero. We find that this scale-independent value agrees well with the linear perturbation theory prediction of the first pairwise velocity moment, $m_{10}$, which is related to the halo \ac{TPCF} as \citep[e.g.,][]{Sheth:2001MNRAS.325.1288S}
\begin{align}
    m_{10}(r) = -\frac{2}{3} \beta \frac{r \bar{\xi}^{\rm R} (r)}{1 + \xi^{\rm R}(r)} \ , \label{eqn:m10_linearPT}
\end{align}
where $\beta \equiv f / b_1$, $f(z)$ is the linear growth rate, $b_1$ is the linear halo bias and $\bar{\xi}^{\rm R}(r)$ is the volume-averaged halo correlation function 
\begin{align}
    \bar{\xi}^{\rm R}(r) \equiv \frac{3}{4 \pi r^3} \int_{0}^{r} 4 \pi \xi^{\rm R}(r') r'^2 \dd{r'} \ .
\end{align}
Accordingly, the relative difference on large scales is approximately given by 
\begin{align}
    \frac{m_{10}^{\rm N1}}{m_{10}^{\rm GR}} - 1 \approx \frac{\beta^{\rm N1}}{\beta^{\rm GR}} - 1 \,.
\end{align}
Note that we have used the fact that the large-scale real-space halo \ac{TPCF} of N1 has been tuned to match the $\Lambda$CDM one. The values of $f(z)$ at $z = 0.5$ and the linear bias $b_1$ are respectively calculated and measured\footnote{To find the linear halo bias $b_1$, we have measured the halo and matter auto-power spectra, taken their ratio and calculated the squart root.} as 
\begin{align}
    f^{\rm GR}(z=0.5) &= 0.761, \quad b_1^{\rm GR} = 1.602 \pm 0.007 \ , \\
    f^{\rm N1}(z=0.5) &= 0.804, \quad b_1^{\rm N1} = 1.527 \pm 0.006 \ ,
\end{align}
which give ${m_{10}^{\rm N1}}/{m_{10}^{\rm GR}} - 1 = 0.108$.
This value (the purple shade region in the lower subpanel of the upper left panel in Fig.~\ref{fig:moments_AllModelsWith10Res_NumDen3p0}) agrees well with what we find in the simulation data (orange line and data points).

In fact, we can qualitatively explain the behaviour of pairwise velocity moments enhancement for $\xi^{\rm R}$-tuned \ac{MG} halo catalogues, for all the 4 velocity moments shown in Fig.~\ref{fig:moments_AllModelsWith10Res_NumDen3p0}, as follows.


The pairwise velocity moments can broadly be thought of as having two contributions: the bulk flow of haloes, which mainly contributes on large scales, and the random motion caused by small-scale shell crossing and virialisation inside dark matter haloes. The pairwise velocity moments can be approximated by the bulk flow and random motion terms, assuming that they are independent of each other so that the cross correlation between them can be ignored. We note that, while the random motions occur on small scales, their contribution still affects the even-order pairwise velocity moments for pairs of haloes at large separations --- this is because random motions do contribute to the velocity difference of the pair, $v$ in Eq.~\eqref{eq:mij_def}, and when taking even powers of $v$ there can be no cancellation and this contribution stays in the final $m_{ij}$.

As mentioned above, the effect of the fifth force in $f(R)$ gravity is suppressed on large scales which are well beyond the range of the force (the inverse of the scalaron Compton wavelength), whereas on small scales we would expect to observe some effect (except where chameleon screening works efficiently to suppress it). On the other hand, in \ac{DGP} models, gravity is enhanced by a constant factor on large scales, but is 
very efficiently suppressed within a few times the typical halo virial radius \citep[see, e.g.,][]{Hernandez-Aguayo:2020kgq}. This implies that in the two \ac{MG} models the contributions from the bulk flow and the random motion will behave very differently.

\begin{figure*}
    \centering 
    \includegraphics[width=0.9\textwidth]{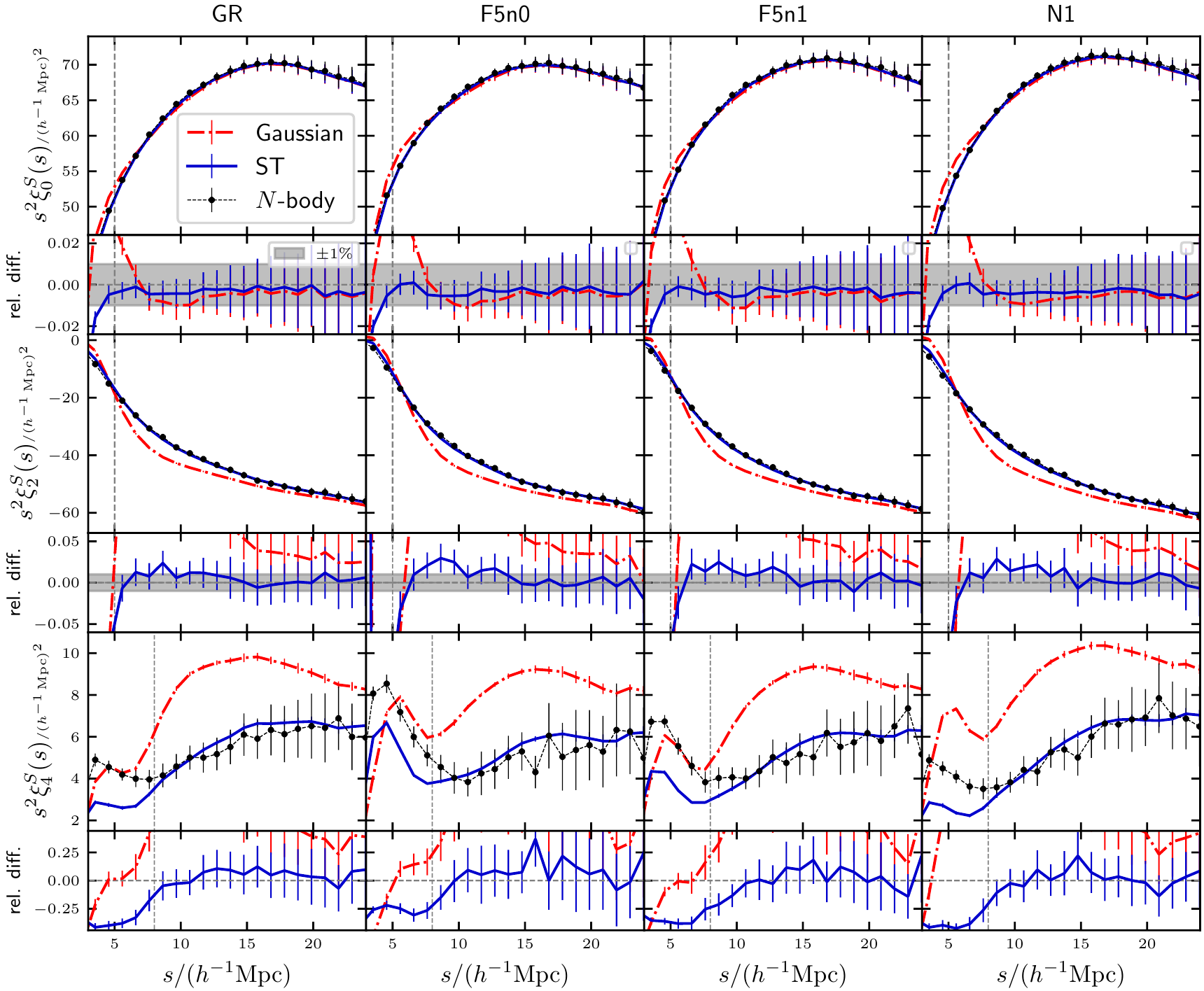}
    \caption{(Colour Online) The monopole, quadrupole and hexadecapole of the redshift-space two-point correlation functions for \ac{GR} (left column), the \ac{DGP} model with $H_0 r_c = 1$ (N1; right column), and $f(R)$ gravity model with $f_{R0} = -10^{-5}$ and $n=0$ (F5n0; second column) and $1$ (F5n1; third column) at $z = 0.5$, from the (\textsc{mg})\textsc{-glam} simulations (black dots). Also shown are the Gaussian (red dash-dotted lines) and ST (blue solid lines) streaming model predictions, where the ingredients of the model are measured from simulations. In the lower sub-panels the relative differences between the \ac{SM} predictions and the simulation measurements, $\xi^{\rm model}(s) / \xi^{\rm sim} (s) - 1$, are shown. The horizontal dashed lines in the lower subpanels denote $0$, the grey shaded regions shows $\pm1\%$ for the monopole and  quadrupole. The vertical dashed lines indicate where the \ac{STSM} predictions start to differ significantly from simulation measurements.
    }
    \label{fig:xiS024_simVSmodel_nd3p0_fsdcsvr}
\end{figure*}

On small scales where the random motions are strong, the velocity moment boost in $f(R)$ gravity can be considerable, since the fifth force is only fully screened in a few very massive haloes, and is unscreened for most objects (at least for the F5n1 and F5n0 models considered here). In DGP models, the Vainshtein screening is efficient on scales smaller than the Vainshtein radius, which causes the \ac{MG} enhancement to be small toward these scales.
This is qualitatively consistent with the small-scale behaviour of $\Delta m_{10} / m_{10}^{\rm GR}$ observed in the upper left subpanel of Fig.~\ref{fig:moments_AllModelsWith10Res_NumDen3p0}.

Linear theory has explained the large-scale behaviour of the first order moment enhancement in both gravity models. Similarly, the leading (linear) term of higher order moment enhancements in perturbation theory, which describes the bulk flow, is given by (see Appendix \ref{sec:linPT_moments} for a heuristic derivation):
\begin{align}\label{eq:c_enhancement_lin}
    \frac{c_n^{\rm MG}}{c_n^{\rm GR}} - 1 \approx \qty(\frac{\beta^{\rm MG}}{\beta^{\rm GR}})^n - 1, \ \text{on large scales}.
\end{align}
The horizontal purple bands in Fig.~\ref{fig:moments_AllModelsWith10Res_NumDen3p0} present the linear predictions for the N1 model. For odd-order pairwise velocity enhancements, the large-scale $N$-body measurements (orange lines) agree well with the linear theory, while for even-order moments, the measurements are systematically smaller. This behaviour can be explained by employing the bulk flow (bf) and random motion (rm) decomposition mentioned above. To be specific, we consider the second-order moment $c_{20}$, but the argument works for any other even-order moments. Consider the large-scale difference of $c_{20}$ between N1 and $\Lambda$CDM and let us decompose the moments into bulk flow and random motion contributions,
\begin{align}
    \Delta c_{20} &\equiv [c]_{\rm MG} - [c]_{\rm GR} \label{eqn:waefdsa1} \\ 
    &= c^{\rm bf}_{\rm MG} + c^{\rm rm}_{\rm MG} - (c^{\rm bf}_{\rm GR} + c^{\rm rm}_{\rm GR}) \label{eqn:waefdsa2}  \\
    &\stackrel{\text{DGP}}{\approx} c^{\rm bf}_{\rm MG} - c^{\rm bf}_{\rm GR}, \label{eqn:waefdsa3} 
\end{align}
where in the last line we have made use of the fact that the random motion contributions in the \ac{DGP} and \ac{GR} models are approximately the same (i.e., $c^{\rm rm}_{\rm DGP}\approx c^{\rm rm}_{\rm GR}$) due to the screening of the fifth force.
Taking the ratio with respect to the moment of $\Lambda$CDM, we have
\begin{align}
    \frac{\Delta c}{c_{\rm GR}} &\approx \frac{c^{\rm bf}_{\rm MG} - c^{\rm bf}_{\rm GR}}{ c^{\rm bf}_{\rm GR} + c^{\rm rm}_{\rm GR}} < \frac{c^{\rm bf}_{\rm MG} - c^{\rm bf}_{\rm GR}}{c^{\rm bf}_{\rm GR}} \approx \qty(\frac{\beta_{\rm MG}}{\beta_{\rm GR}})^2 - 1 \,, \label{eqn:waefdsa4}
\end{align}
where for the inequality we have implicitly used the facts $c^{\rm bf}_{\rm DGP}>c^{\rm bf}_{\rm GR}$ and $c^{\rm rm}_{\rm GR}>0$ (as mentioned above, for even moments, $c^{\rm rm}_{\rm GR}\ne0$ even for halo pairs at large separations). This is the reason why the even-order moments of the simulation measurements are lower than those predicted by the linear theory, Eq.~\eqref{eq:c_enhancement_lin}.

\begin{figure*}
    \centering 
    \includegraphics[width=0.9\textwidth]{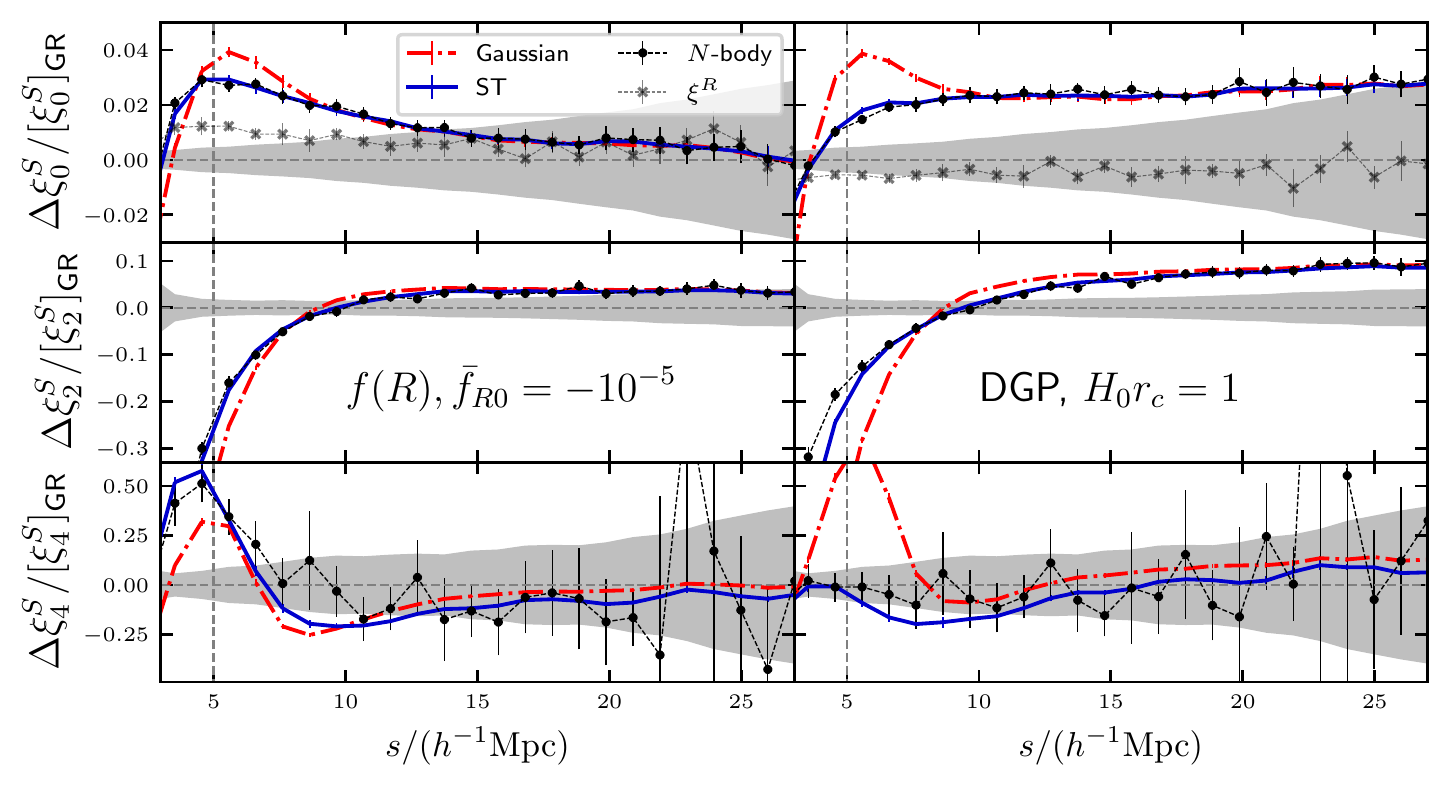}
    \caption{(Colour Online) The relative difference of redshift-space two-point correlation function multipoles between F5n1 and \ac{GR} (left column), and between N1 and \ac{GR} (right column), from the (\textsc{mg})\textsc{-glam} simulations. The black dots with error bars are the measurements from simulations, and the small crosses with error bars in the top row are the relative differences in the real-space halo \ac{TPCF}s --- these are measured from the tuned halo catalogues and are there just for comparison. The streaming model predictions are plotted as the red dash-dotted (\ac{GSM}) and blue solid (\ac{STSM}) lines. The gray-shaded areas correspond to the standard deviation for \ac{GR} over $72 \times 3$ measurements obtained from $72$ realisations of the \textsc{glam} code. }
    \label{fig:MGGLAM_xiS024_ratio_MGvsGR_F5n1_nd3p0_ewfdvsr}
\end{figure*}

For the odd-order moments, Eqs.~\eqref{eqn:waefdsa1}-\eqref{eqn:waefdsa3} are still valid. The difference arises in the inequality of Eq.~\eqref{eqn:waefdsa4}. Random motion contributions to odd-order velocity moments are significantly smaller than those of even-order moments, since even-order powers of $v$ are always positive and therefore can have no cancellation in Eq.~\eqref{eq:mij_def}, but odd-order powers of $v$ in Eq.~\eqref{eq:mij_def} can so that $c^{\rm rm}_{\rm GR}\approx0$. We therefore have, for odd-order moments, 
\begin{align}
    \frac{\Delta c}{c_{\rm GR}} &\approx \frac{c^{\rm bf}_{\rm MG} - c^{\rm bf}_{\rm GR}}{ c^{\rm bf}_{\rm GR} + c^{\rm rm}_{\rm GR}} \approx \frac{c^{\rm bf}_{\rm MG}- c^{\rm bf}_{\rm GR}}{ c^{\rm bf}_{\rm GR} } \approx \qty(\frac{\beta_{\rm MG}}{\beta_{\rm GR}})^2 - 1 \ . \label{eqn:waefdsa5}
\end{align}

\subsubsection{Halo real space two-point correlation functions}

As mentioned above, we have constructed the \ac{GR} halo catalogues to have a fixed number density, and tuned the minimum mass cut $M_{\rm min}$ of the various \ac{MG} halo catalogues so that the halo real-space TPCFs match their \ac{GR} counterparts closely. 
As shown in the right panel of Fig.~\ref{fig:cHMF_DiffSim_comp}, the correlation functions of the tuned $f(R)$ gravity and \ac{DGP} halo catalogues agree with the $\Lambda$CDM counterparts on scales larger than $\sim 10 \, h^{-1}\mathrm{Mpc}$ to within a relative difference  of less than $1\simeq$ per cent.

The effect of \ac{MG} on the \ac{TPCF} of halo catalogues with fixed $n_h$, or fixed minimum mass cut $M_{\rm min}$, is interesting in its own right. However, this has been investigated in various previous works, more recently by \citet{2020arXiv201105771A}. On the other hand, as we have discussed above, the tuning of $n_h$ to make it \ac{MG} model dependent---in order to achieve a matching of the real-space halo \ac{TPCF}s in  different models---leads to catalogues where the effect of different halo populations can be more cleanly separated from that of the fifth force. In addition, since the real-space halo \ac{TPCF}s are matched, any difference in the redshift-space clustering is necessarily caused by the difference in the pairwise velocities. This makes the interpretation of the underlying physics more straightforward. As a result, for the rest of this paper we will only use the tuned halo catalogues.

\subsection{Halo redshift-space two-point correlation function multipoles}
\label{subsec:halo_2pcf_res}

In this subsection, we will apply the \ac{ST} velocity distribution to the streaming model, to predict halo redshift-space correlation function monopoles, quadrupoles and hexadecapoles.
Since our goal is to show that the \ac{ST} model is generic and applicable to both $\Lambda$CDM and \ac{MG} models, we measure all the ingredients of the streaming model, including halo real-space correlation functions and pairwise velocity moments, from the simulations. We use the \textsc{mg-glam} simulation data for this investigation. The halo catalogues at $z = 0.5$ with the halo number density around $10^{-3}~(h^{-1} \mathrm{Mpc})^{-3}$ are used in this section. 
We measure the model ingredients for each realisation, compute the streaming model predictions and then present the average and standard deviation.
The results with other number densities and in other redshifts are qualitatively similar, and some of these will be shown in Appendix~\ref{appendix:more_cases}.

In Fig.~\ref{fig:xiS024_simVSmodel_nd3p0_fsdcsvr}, we show the multipoles of the redshift-space correlation function, $\xi^S_{0,2,4} (s)$ (the different rows), for the $\Lambda$CDM, F5n0, F5n1 and N1 models (the different columns). In the lower subpanels of each panel, the relative differences between the model predictions and the simulation measurements are displayed. The monopole predictions are quite accurate for both \ac{GSM} and \ac{STSM}, while it is apparent that \ac{STSM} performs slightly better. If one targets at percent-level accuracy, then \ac{GSM} fails at $s \lesssim 10 \, h^{-1}\mathrm{Mpc}$ while \ac{STSM} works well down to $s\simeq 5 \,h^{-1}\mathrm{Mpc}$.
The improvement made by using the ST velocity distribution is significant when we consider the quadrupole.
The \ac{GSM} is biased for scales $\lesssim 20 \, h^{-1}\mathrm{Mpc}$, whilst the \ac{STSM} only starts failing on scales smaller than $5 \, h^{-1}\mathrm{Mpc}$. Similar behaviour is found for the hexadecapole.
Although the measurement of the hexadecapole is rather noisy, mainly due to the small simulation box size, the ST model agrees with simulation measurements within one standard deviation for scales larger than $\simeq 8 \, h^{-1}\mathrm{Mpc}$, while the \ac{GSM} is biased on all scales considered here.


The \ac{STSM} matches the $\xi^{\rm S}_{0,2,4}(s)$ on scales larger than the maximum separation shown in Fig.~\ref{fig:xiS024_simVSmodel_nd3p0_fsdcsvr}, so to improve the readability of the figure, we opt not to show the behaviour on larger scales. We conclude that the ST pairwise velocity distribution with the streaming model is competent in predicting redshift-space correlation functions in $f(R)$ gravity and the \ac{DGP} model, as well as in \ac{GR}.

Note that this excellent performance of the \ac{ST} model is under ideal conditions: all ingredients of the streaming model of RSD, e.g., the halo real-space two-point correlation function, $\xi^{\rm R}(r)$, and the four lowest order pairwise velocity moments, are all measured from simulations, instead of using theoretical models. 
We will briefly discuss our plan on constructing simulation-based emulators for $\xi^{\rm R} (r)$ and higher-order pairwise velocity moments in Section~\ref{sec:conc_and_disc} to extend the unbiased predictions down to highly non-linear scales.

Fig.~\ref{fig:MGGLAM_xiS024_ratio_MGvsGR_F5n1_nd3p0_ewfdvsr} compares the enhancements, with respect to $\Lambda$CDM, of the measured RSD monopole (upper panels), quadrupole (middle) and hexadecapoe (lower) from the simulated halo catalouges (symbols with error bars), against the predictions by the Gaussian (red dashed lines) and ST (blue solid) streaming models. The left column shows the results for F5n1 and the right panel for N1. We note that, again, for both MG models, STSM outperforms the GSM in matching the simulation data. For the monopole, GSM starts to fail at $\sim 10 \, h^{-1}\mathrm{Mpc}$ while STSM works well down to $\sim 3 \, h^{-1}\mathrm{Mpc}$. For the quadrupole, the GSM prediction deviates from simulation data at $s \lesssim 20 \, h^{-1}\mathrm{Mpc}$, while STSM remains in good agreement with the latter down to $\sim 5 \, h^{-1}\mathrm{Mpc}$. For hexadecapole, we can see an improvement in STSM as well, though here the simulation data is noisier.

\subsection{Schematic demonstration of scale dependence of the constraint on \ac{MG} parameters}
\label{subsec:fisher}

We have seen that, compared with the traditional Gaussian model, the ST model has achieved greater success in predicting halo clustering on smaller scales ($5 \text{-} 25 \, h^{-1}\mathrm{Mpc}$). In order to quantitatively demonstrate the constraining power gained from small-scale RSD signals, we will forecast the constraints on the \ac{MG} parameters using a highly simplified Fisher analysis, in which all parameters are fixed except the \ac{MG} parameters such $\bar{f}_{R0}$ in  $f(R)$ gravity and $H_0 r_c$ in \ac{DGP}.

The Fisher matrix method provides a way to propagate the observable uncertainty to the constraints of cosmological parameters. Our calculation of the Fisher matrix is based on \citet{1997PhRvL..79.3806T} and \citet{2003ApJ...598..720S}, assuming a Gaussian likelihood function for our measurements of the correlation function multipoles.
Additionally we ignore any parameter dependence of the covariance matrix, in which case the Fisher matrix of a redshift slice centered at $z$ can be approximated as 
\begin{align}
    F_{ij} &= \sum_{\alpha \beta} \frac{\partial f_\alpha}{\partial p_i} \mathrm{Cov}^{-1} [f_\alpha, f_\beta] \frac{\partial f_\beta}{\partial p_j} \label{eqn:fishermatrix}
\end{align}
where Greek indices $\alpha, \beta$ label the spatial separation bins, e.g., $s_\alpha$; $f_\alpha = \{\xi^{\rm S}_{0} (s_\alpha), \xi^{\rm S}_{2} (s_\alpha) \}$ are the redshift-space halo correlation function multipoles at redshift $z$; $\mathrm{Cov}[f_\alpha, f_{\beta}]$ is the corresponding covariance matrix and $p_i, p_j$ are the model parameters (only one parameter in our simplified case here) being considered, which are $\{\bar{f}_{R0} \}$ in $f(R)$ gravity and $\{ H_0 r_c \}$ in the DGP model. The covariance matrix of the redshift-space multipole moments are calculated from the halo catalogues of $72$ \textsc{glam} $\Lambda$CDM runs. The $1\sigma$ error is given by $\sqrt{(F^{-1})_{11}}$. We fix the maximum separation $s_{\rm max}=35 \, h^{-1}\mathrm{Mpc}$, and vary the minimum scale $s_{\rm min}$ from $22$ to $2.5 \, h^{-1} \mathrm{Mpc}$ to explore the constraining power on \ac{MG} parameters gained from small-scale information.

For the F5n1 and N1 models considered here, the derivatives in Eq.~\eqref{eqn:fishermatrix} are approximated by (taking $\xi^{\rm S}_0 (s)$ as an example)
\begin{align}
    \frac{\partial \xi^{\rm S}_0 (s_\alpha; \bar{f}_{R0})}{\partial \bar{f}_{R0} } \approx \frac{\xi^{\rm S}_0 (s_\alpha; -10^{-5}) - \xi^{\rm S}_0 (s_\alpha; \text{GR}) }{-10^{-5}} \\
\intertext{and}
    \frac{\partial \xi^{\rm S}_0 (s_\alpha; \gamma)}{\partial (H_0 r_c)} \approx \frac{\xi^{\rm S}_0 (s_\alpha; 1) - \xi^{\rm S}_0 (s_\alpha; \gamma_{\text{GR}}=0) }{1} ,
\end{align}
respectively, where $\gamma \equiv 1/(H_0 r_c)$.

\begin{figure*}
    \centering 
    \includegraphics[width=0.9\textwidth]{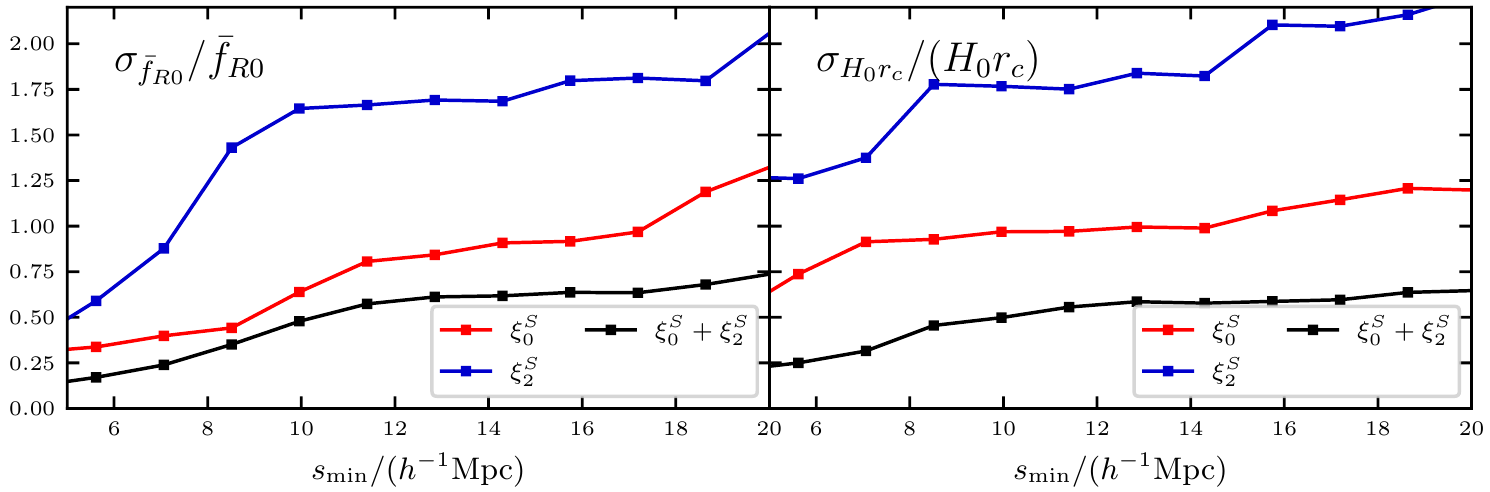}
    \caption{(Colour Online) The dependence of the constraining power on $\bar{f}_{R0}$ (left) and $H_0 r_c$ (right) from halo redshift-space correlation function multipole measurements, on the minimum scale $s_{\rm min}$ included in the constraints, based on our simplified Fisher forecast with different minimum scales considered. The maximum scale is fixed to $s_{\rm max} = 35 \, h^{-1}\mathrm{Mpc}$. Three scenarios are considered: monopole data alone (red), quadrupole data only (blue), and including both monopole and quadrupole (black).
    }
    \label{fig:deltafR_ff}
\end{figure*}

The Fisher forecast results are presented in Fig.~\ref{fig:deltafR_ff}, in which we have considered three scenarios: using monoopole data only (red), using quadrupole data only (blue) and using both monopole and quadrupole (black). In all cases it is clear that the constraining power on both $\bar{f}_{R0}$ and $H_0 r_c$ monotonically increases with decreasing $s_{\rm min}$.
Compared with using the monopole data alone, the addition of the quadrupole data (which on its own does not produce very strong constraints) tightens the constraints by $\sim 20\%$. Including the hexadecapoles leads to little improvement, which is unsurprising given the rather noisy hexadecapole measurements. Most interestingly, we note that, for both $f(R)$ gravity and \ac{DGP}, including scales of $s\lesssim 10 \, h^{-1}\mathrm{Mpc}$ can markedly improve the constraints on the \ac{MG} parameter. This confirms that small-scale RSD, if measured precisely and modelled accurately, can be a promising tool to help test gravity models using galaxy clustering data.

\section{conclusions and discussion}
\label{sec:conc_and_disc}

Percent-level accuracy in modeling the anisotropies of redshift-space galaxy clustering is required to accurately recover cosmological information from \ac{RSD} signals in order to distinguish between dark energy and modified gravity scenarios. Within the framework of the streaming model of \ac{RSD}, this requires that, compared with the current status, we must: (i) improve the mapping of real- to redshift-space correlations, i.e., find a better description of the pairwise velocity distribution, (ii) increase the accuracy of the predictions of the streaming model ingredients, including the halo real-space correlation function $\xi^{\rm R}(r)$ and the pairwise velocity moments. We have investigated both aspects in the context of modified gravity cosmologies. For the first aspect, we have demonstrated that the \ac{ST} probability distribution for the halo pairwise velocity, which was introduced by \citet{Cuesta-Lazaro:2020MNRAS.498.1175C} as an alternative to the traditional Gaussian model, is applicable to the measurements from the $N$-body simulations of all \ac{MG} models considered here. For the second aspect, we have explored \ac{MG} signals in both the individual ingredients and predictions of the streaming model. This work has made full use of \textsc{mg-glam} \citep{Hernandez-Aguayo:MG-GLAM2021,Ruan:2021MGGLAMfR}, a new code for fast production of full $N$-body simulations in a wide range of \ac{MG} models.

The traditional Gaussian distribution fails to fully capture some properties of the halo pairwise velocity \ac{PDF} found in $N$-body simulations, such as the skewness and kurtosis. The \ac{ST} distribution can be tuned to match the four lowest order velocity moments of $\mathcal{P} (v_{\parallel} | \bm{r})$ with four parameters (two more than a Gaussian). Compared with the Gaussian form, the \ac{ST} model extends the validity of the streaming model from $\simeq 7$ to $\lesssim 5 \, h^{-1} \mathrm{Mpc}$ for the monopole, and $\simeq15$ to $\simeq5 \, h^{-1} \mathrm{Mpc}$ for the quadrupole. For the hexadecapole, the \ac{ST} model gives predictions that are correct down to about $8 \, h^{-1} \mathrm{Mpc}$, while the \ac{GSM} is biased on all scales shown. The performance of \ac{ST} is equally good among all considered gravity models, including $\Lambda$CDM, $f(R)$ gravity with $\bar{f}_{R0} = -10^{-5}, n = 0$ (F5n0) and $1$ (F5n1), and the normal branch of \ac{DGP} with $H_0 r_c = 1$ (N1).

We have investigated the \ac{MG} enhancements of halo pairwise velocity moments and redshift-space correlation functions with respect to the $\Lambda$CDM baseline. To remove the effect of different halo populations on large scales and make the physics in the results easier to interpret, we have tuned the mass cut of \ac{MG} halo catalogues to match the real-space correlation functions to that of the $\Lambda$CDM halo catalogues on large scales. With this, the \ac{MG} pairwise velocity moment enhancements on large scales can be explained by linear theory.

We have performed a simple Fisher forecast analysis to assess the impact of including small-scale information on the power of \ac{RSD} in testing and distinguishing different gravity models. Fig.~\ref{fig:deltafR_ff} demonstrates that both including the \ac{RSD} quadrupole and including data from scales of $s\lesssim15 \, h^{-1}\mathrm{Mpc}$ can substantially decease the uncertainty in the constrained \ac{MG} parameters. This highlights the potentially important role played by small-scale \ac{RSD} in cosmological tests of gravity using data from upcoming galaxy surveys such as DESI and Euclid.

The analysis in this work has been largely theoretical, since we have focused on haloes and made use of direct measurements from simulations for a fixed number of theoretical models. To apply the \ac{ST} model to real galaxy survey data and fully exploit its accuracy on small scales, we need to improve in a couple of aspects. First of all, we need accurate predictions of the two \ac{SM} model ingredients---the pairwise velocity moments and real-space correlation functions---for arbitrary cosmological models. While on linear and quasi-linear scales, perturbation based approaches have proven very useful in this regard, since our focus here is on the small, non-linear scales, where the perturbative approach fails, alternatives need to be sought. We plan to build emulators for the pairwise velocity PDF moments and \ac{TPCF}s, taking advantage of the large number of big simulations that will be enabled by the fast \textsc{mg-glam} code. 
Note that this is different from directly emulating the 2D redshift-space correlation functions or their multipoles, since the \ac{SM} ingredients are quantities with clearer physical meanings.

Secondly, we need to extend our analysis to observable tracers of the large-scale structure, such as galaxies. The inclusion of satellite galaxies in clustering analysis will lead to a substantial finger-of-God effect, which can pose new challenges to the accurate modelling of small-scale RSD, and this needs to be investigated. Also, since our simulations are dark matter only, a model of galaxy-halo connection, such as \ac{HOD} \citep[e.g.,][]{Berlind:2002rn,Zheng:2004id}, abundance matching \citep[e.g.,][]{Conroy:2005aq,Moster:2010,Reddick:2013}, and semi-analytic galaxy formation models \citep[e.g.,][]{1993MNRAS.264..201K,1994MNRAS.271..781C,2016MNRAS.462.3854L}, is needed, and this will likely introduce additional uncertainties in the predicted signal. 
For simulations with relatively low resolution, the \ac{HOD} method is usually adopted to construct galaxy mock catalogues by populating the simulated haloes, where the \ac{HOD} parameters can be calibrated to match the observed galaxy number density and certain properties of their large-scale correlation. 
In a forthcoming project, we will focus on the redshift-space \ac{TPCF} multipoles for \ac{HOD} galaxies, and use these to reassess the constraining power of small-scale \ac{RSD} in gravity tests.
Modelling the redshift-space galaxy clustering within the streaming model gives us more flexibility when combining with the \ac{HOD} prescription.

\section*{Acknowledgements}
We wish to thank Christian Arnold for kindly providing the MG \textsc{lightcone} simulation data \citep{2019MNRAS.483..790A}.
C-ZR, AE and BL are supported by the European Research Council (ERC) through a starting Grant (ERC-StG-716532 PUNCA). BL and CMB are further supported by the UK Science and Technology Funding Council (STFC) Consolidated Grant No.~ST/I00162X/1 and ST/P000541/1. CH-A acknowledges support from the Excellence Cluster ORIGINS which is funded by the Deutsche Forschungsgemeinschaft (DFG, German Research Foundation) under Germany's Excellence Strategy - EXC-2094-390783311. FP thanks the support of the Spanish Ministry of Science and Innovation funding grant PGC2018- 101931-B-I00.

This work used the DiRAC@Durham facility managed by the Institute for Computational Cosmology on behalf of the STFC DiRAC HPC Facility (\url{www.dirac.ac.uk}). The equipment was funded by BEIS via STFC capital grants ST/K00042X/1, ST/P002293/1, ST/R002371/1 and ST/S002502/1, Durham University and STFC operation grant ST/R000832/1. DiRAC is part of the UK National e-Infrastructure.

This work used the skun6@IAA facility (\url{www.skiesanduniverses.org}) managed by the Instituto de Astrof\'{i}sica de Andaluc\'{i}a (CSIC). The equipment was funded by the Spanish Ministry of Science EU-FEDER infrastructure grants EQC2018-004366-P and EQC2019-006089-P.

\section*{Data Availability}

The data underlying this article will be shared on reasonable request to the corresponding author. An example code used for the numerical integrals in the ST streaming model calculation is shared \href{https://github.com/chzruan/StreamingModelExample}{here}; see also Appendix \ref{appendix:details}.




\bibliographystyle{mnras}
\bibliography{main.bib} 



\appendix

\section{Method of moments for the ST distribution}
\label{appendix:app1}

The four parameters of the skew-T distribution, $v_c, w, \alpha, \nu$, can be analytically related to the first four moments. To simplify the relation between these moments and parameters, let us introduce
\begin{equation}
\begin{split}
b_\nu &= \left( \frac{\nu}{\pi} \right)^\frac{1}{2} \frac{\Gamma \left( \frac{\nu - 1}{2} \right)}{\Gamma(\nu/2)},\\
\delta &= \frac{\alpha}{\sqrt{(1 + \alpha^2)}},\\
\gamma_1 &= \frac{c_3}{c_2^{3/2}},\\
\gamma_2 &= \frac{c_4}{c_2^2}.
\end{split}
\end{equation}
The moments can then be written as,
\begin{equation}
m_1 = v_c + w \delta b_\nu, \label{eqn:st_m1_moments}
\end{equation}
\begin{equation}
c_2 = w^2 \left[ \frac{\nu}{\nu - 2} - \delta^2 b_\nu^2 \right], \label{eqn:st_c2_moments}
\end{equation}
\begin{equation}
\gamma_1 = \delta b_\nu 
\left[ \frac{\nu ( 3 - \delta^2)}{\nu - 3} - \frac{3 \nu}{\nu -2} + 2 \delta^2 b_\nu^2  \right] \left[\frac{\nu}{\nu-2} - \delta^2 b_\nu^2\right]^{-\frac{3}{2}}, \label{eqn:st_g1_moments}
\end{equation}
\begin{equation}
\begin{split}
\gamma_2 = &\left[ \frac{3 \nu^2}{(\nu -2)(\nu-4)} 
					- \frac{4 \delta^2 b_\nu^2 \nu (3 - \delta^2)}{\nu - 3}
                    - \frac{6 \delta^2 b_\nu^2 \nu}{\nu - 2} - 3 \delta^4 b_\nu^4 \right]\\
                    & \left[\frac{\nu}{\nu -2} - \delta^2 b_\nu^2 \right]^{-2} .
 \end{split} \label{eqn:st_g2_moments}
\end{equation}
These form a system of nonlinearly coupled algebraic equations that can be solved numerically: parameters $w,\alpha$ and $\nu$ are obtained from the last three equations given the variance, skewness and kurtosis of the distribution, and the remaining parameter, $v_c$, can then directly be obtained from the equation for the mean.

\section{Convergence Tests}
\label{appendix:ConvergenceTests}

The \textsc{glam} and \textsc{mg}-\textsc{glam} simulations used in this work have a relatively small box size, $512\,h^{-1}\mathrm{Mpc}$. While their mass resolution is high compared to many other \ac{MG} simulations to date, the particle-mesh nature of the \textsc{glam}-based codes means that the force resolution in these runs is poorer than what could be achieved using adaptive-mesh-refinement simulations with the same particle number and box size. In addition, \textsc{mg}-\textsc{glam} is a relatively new code and, while it has passed various tests as demonstrated in the code papers, those tests do not include velocity field statistics such as the ones considered in this work. For these reasons, in this Appendix we will carry out a test of the latter, by comparing the first four halo pairwise velocity moments measured from \textsc{mg}-\textsc{glam} simulations of F5n1 and GR, with the predictions by a higher-resolution \textsc{lightcone} simulation for the same models.

The \ac{MG} \textsc{lightcone} simulation project \citep{2019MNRAS.483..790A} provides a set of high-resolution cosmological simulations of GR and F5n1, which employs the \ac{MG} $N$-body code \textsc{mg-gadget} \citep{2013MNRAS.436..348P}, adopting the same Planck15 $\Lambda$CDM cosmology. Starting from identical initial conditions, the pair of GR and F5n1 simulation runs followed the dynamical evolution of $2048^3$ dark matter particles in a box with $768 \, h^{-1}\mathrm{Mpc}$ comoving length, reaching a mass resolution of $M_{\text{particle}} = 4.50 \times 10^9 \, h^{-1} M_{\odot}$. 
This high mass resolution make sure that the measured halo clustering signals are precise on small scales (down to $\sim 1 \, h^{-1} \mathrm{Mpc}$).
The halo catalogues are obtained with the \textsc{subfind} \citep{springel2001populating} algorithm. 
The halo mass definition adopted is $M_{\rm 200c} \equiv \frac{4\pi}{3} (r_{\rm 200c})^3 200 \rho_{\rm c}$, where $\rho_{\rm c} \equiv 3H^2 / (8\pi G)$ is the critical density of the Universe, and $r_{\rm 200c}$ is the spherical halo radius within which the spherically averaged mass density equals $200$ times $\rho_{\rm c}$. The halo catalogues at redshifts $z=1$ and $0$ are available.

\begin{figure*}
    \centering
    \includegraphics[width=0.9\textwidth]{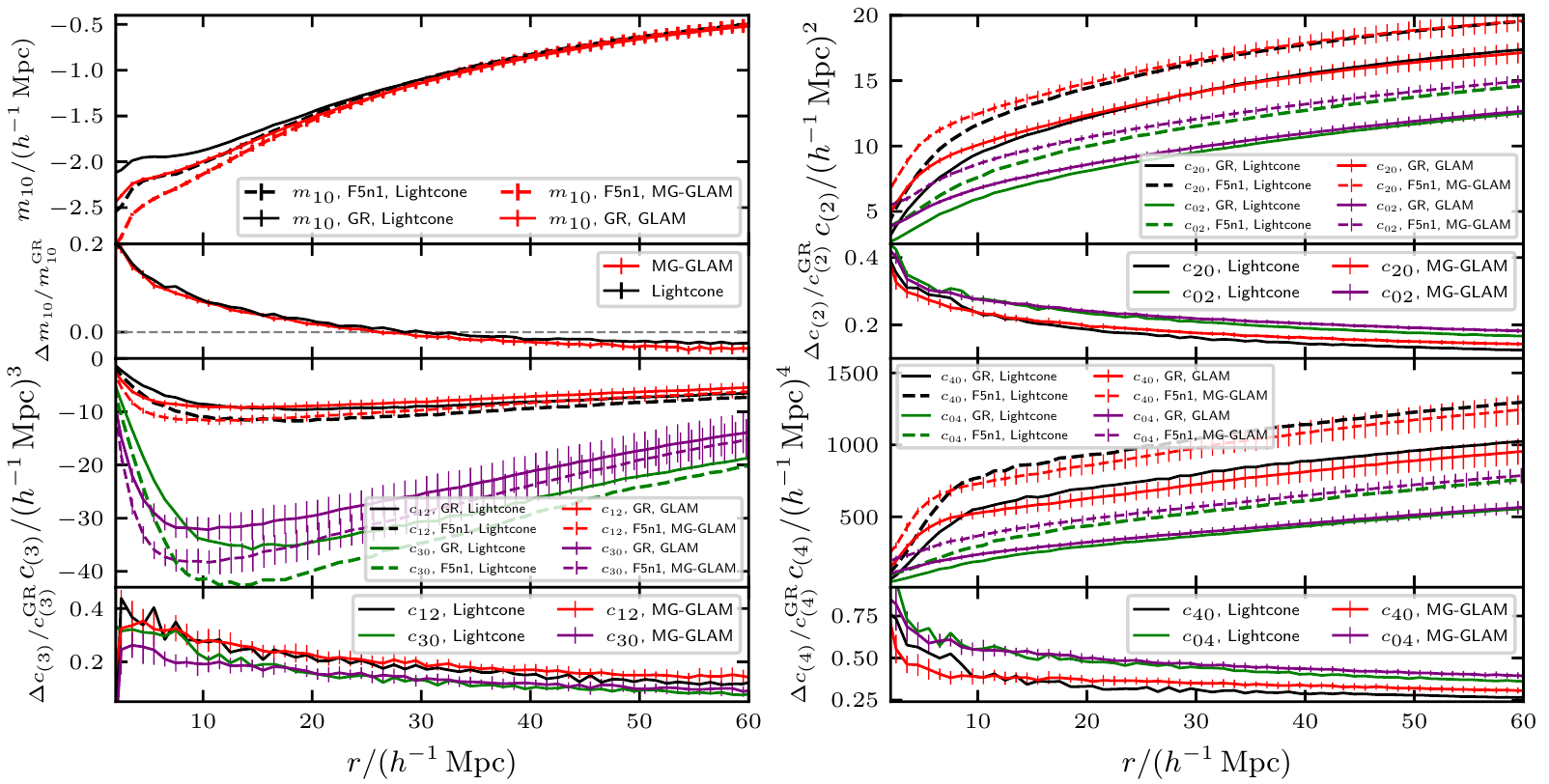}
    \caption{The four lowest order moments of the radial and transverse halo pairwise velocity for the \ac{GR} and F5n1 models at $z = 0$, from the MG light-cone project and \textsc{mg-glam} simulations. The lower subpanels show the relative difference between the velocity moments of the F5n1 and \ac{GR}.}
    \label{fig:moments_MGGLAMvsLC_erwsfdcv}
\end{figure*}

The results are shown in Fig.~\ref{fig:moments_MGGLAMvsLC_erwsfdcv}, where the four panels from upper left to lower right are respectively for the first-, second-, third- and fourth-order moments. In each panel, the upper subpanel shows the magnitudes of the moments as a function of the halo separation $r$, while the lower subpanel shows the relative difference between F5n1 and GR. The line styles and colours are indicated by legends. All results are at $z=0$.

When reading Fig.~\ref{fig:moments_MGGLAMvsLC_erwsfdcv}, let us bear in mind that the halo populations in the (\textsc{mg})-\textsc{glam} and \ac{MG} \textsc{lightcone} simulations are necessarily different. 
This suggests that these two sets of simulations can have different halo biases, which would affect the amplitudes of the velocity moments, although the shapes are broadly the same, as can be readily seen. We have not made attempts to force an agreement by adjusting the halo number densities in the different simulations, but we have checked this using several $\Lambda$CDM simulations that use different simulation codes, resolutions and halo finders, and found similar levels of discrepancy among all of them.

More interestingly, the lower subpanels show that the model difference between F5n1 and GR predicted by the two sets of simulations agree very well for all the analysed velocity moments, and all halo separations shown in Fig.~\ref{fig:moments_MGGLAMvsLC_erwsfdcv}. In particular, the first moment, $m_{10}$, which the redshift-space halo correlation function quadrupole is most sensitive to, agrees between the two sets of simulations down to $\sim 5 \, h^{-1}$Mpc. This shows that the \textsc{mg}-\textsc{glam} simulation results can be used to study RSD at such small scales.

Fig.~\ref{fig:moments_MGGLAMvsLC_erwsfdcv} also shows that the enhancements of the velocity moments are different on small scales between the \textsc{mg-glam} and \textsc{lightcone} simulations.
This difference is in most cases at a few percent at most above $\simeq 10 \,h^{-1}\textrm{Mpc}$, and --- in the case of $m_{10}$ --- around the percent level.
This level of difference is expected given the many differences in the two sets of simulations, from codes to technical specifications, such as details of halo finding. 
The effect of these simulation/technical differences can also be seen by comparing the values (rather than the model differences) of the moments predicted by the two simulations (the upper subpanels of each panel). 

At $< 10 \,h^{-1}\textrm{Mpc}$, we see a stronger discrepancy between the two simulations for $c_{30}$ and $c_{40}$, at $\simeq 10\%$ (for $c_{04}$ the agreement is much better --- see the green/purple curves in the lower right panel of Fig.~\ref{fig:moments_MGGLAMvsLC_erwsfdcv}). 
Again, given the more significant differences between the absolute curves of $c_{30}$ and $c_{40}$ for the same models, such as GR, as shown in Fig.~\ref{fig:moments_MGGLAMvsLC_erwsfdcv} (the green solid and purple solid curves in the upper subpanel of the lower left panel), this is not surprising. Note in particular that \textsc{mg-glam} uses $M_{\rm vir}$ while \textsc{lightcone} uses $M_{200c}$ as the halo mass definition.

\section{The performance of the \ac{ST} model in more models}
\label{appendix:more_cases}

In Fig.~\ref{fig:xiS024_simVSmodel_nd3p0_fsdcsvr}, we have demonstrated that the ST streaming model works very well in predicting the redshift-space correlation function multipoles $\xi^{\rm S}_{0,2,4}$ in not only the GR model, as found by \cite{Cuesta-Lazaro:2020MNRAS.498.1175C}, but also for several modified gravity models. However, due to space limit, in that figure we have only presented the results at a single redshift ($z=0.5$) and around a single halo number density $n_h = 10^{-3.0} \, (h^{-1}{\rm Mpc})^{-3}$. 

\begin{figure*}
    \centering 
    \includegraphics[width=0.8\textwidth]{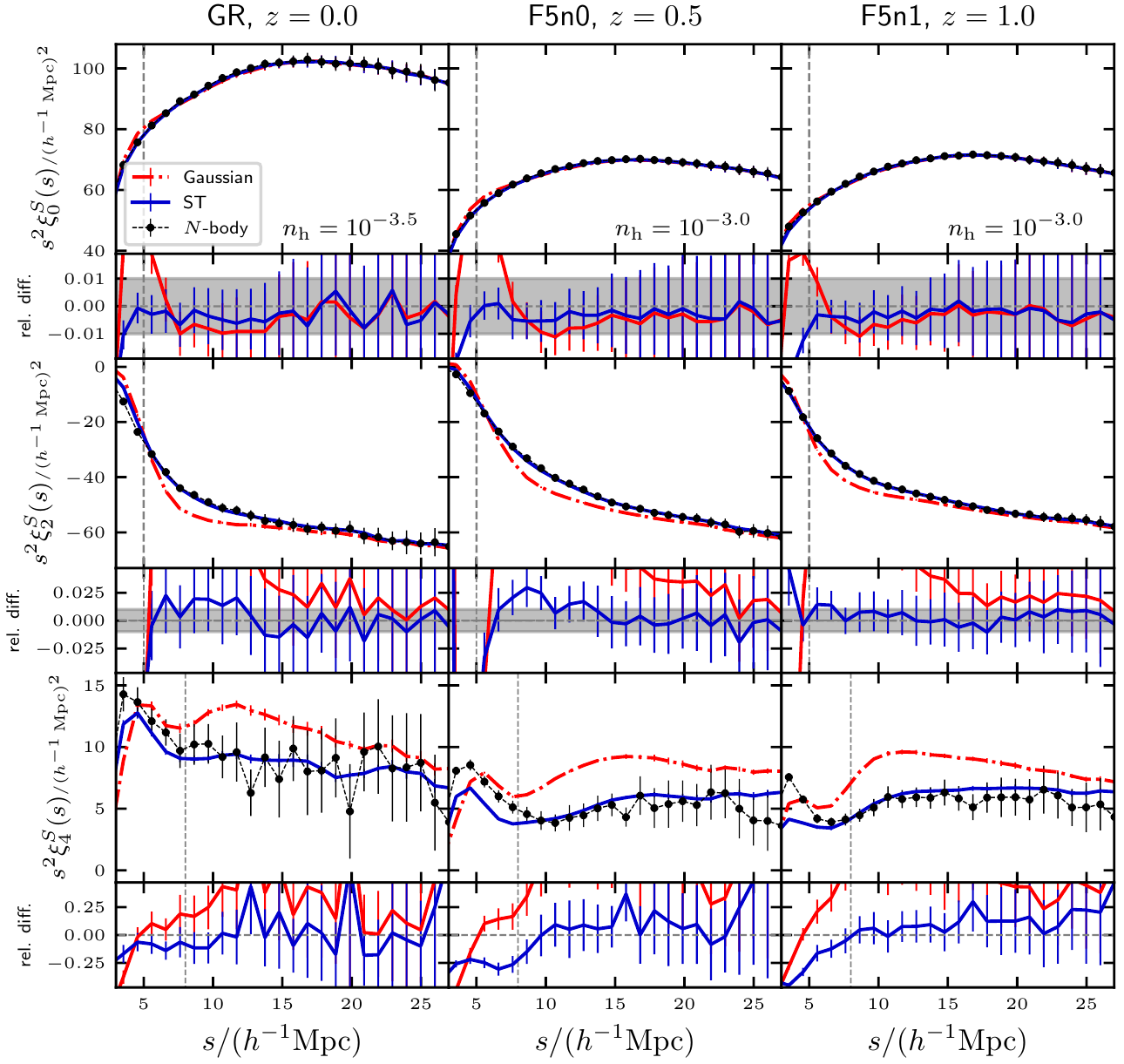}
    \caption{(Colour Online) The monopole, quadrupole and hexadecapole of the redshift-space two-point correlation functions for \ac{GR} (left column), $f(R)$ gravity model with $f_{R0} = -10^{-5}$ and $n=0$ (F5n0; middle column) and $1$ (F5n1; right column), at different redshifts as indicated by the subtitles, from the \textsc{mg-glam} simulations  (black dots), along with the Gaussian (red lines) and ST (blue lines) streaming model predictions. 
    In the lower sub-panels the relative differences between the \ac{SM} predictions and the simulation measurements, $\xi^{\rm model}(s) / \xi^{\rm sim} (s) - 1$, are shown. The horizontal dashed lines in the lower subpanels denote $0$, the grey shaded regions denoting $\pm 1\%$ for the monopoles and quadrupoles. The vertical dashed lines indicate where the \ac{STSM} predictions start to differ significantly from simulation measurements.
    }
    \label{fig:xiS024_simVSmodel_morecases_fsdcsvr}
\end{figure*}

We have also carried out similar checks for a range of other redshifts and halo number densities, and in all cases we found similarly good agreement between the ST streaming model and simulation predictions of RSD multipoles. A few selected examples are shown in Fig.~\ref{fig:xiS024_simVSmodel_morecases_fsdcsvr}. The left, middle and right columns are respectively GR $n_h = 10^{-3.5} \left(h^{-1}{\rm Mpc}\right)^{-3}$ at $z = 0$, F5n1 $n_h=10^{-3.0} \left(h^{-1}{\rm Mpc}\right)^{-3}$ at $z=0.5$, and F5n1 $n_h=10^{-3} \left(h^{-1}{\rm Mpc}\right)^{-3}$ at $z=1$. The three rows are for $\xi^{\rm S}_{0,2,4}$ respectively. In each panel, the upper subpanel compares simulation measurement (symbols with error bars) with the predictions of the Gaussian (red) and ST (blue) streaming models, and the lower subpanel shows the relative differences between the two streaming models with respect to the simulation measurement. In all the cases, the ST streaming model clearly gives more reliable predictions than the Gaussian one, indicating that the former can be applied to the modified gravity models studied in this work. There is no apparent reason why we should not expect it to work for other models as well. The performance of the \ac{GSM} is better in the lower halo number density case, which is expected, since the pairwise velocity \ac{PDF} becomes more Gaussian for more massive haloes.

Although not shown here, we have also checked the ST streaming model for a few other $\Lambda$CDM simulations which were run using different codes, at different resolutions and with different halo finders. In all cases the agreement with simulation prediction is equally good.

\section{Linear perturbation predictions of halo pairwise velocity moments}
\label{sec:linPT_moments}

In this appendix we aim to present a derivation of Eq.~\eqref{eq:c_enhancement_lin}, showing that in \textit{linear} perturbation theory the $n$-th order halo pairwise velocity (central) moments, $c_n$, scale as $\beta^n$, where $\beta \equiv f/b_1$ was introduced in Eq.~\eqref{eqn:m10_linearPT}. 

We start from the line-of-sight peculiar velocity difference between points $\bm{x}$ and $\bm{x}'$, expressed in terms of the velocity divergence $\theta(\bm{x}) \equiv \bm{\nabla} \cdot \bm{v}(\bm{x})$,
\begin{align}
    v_{\parallel} &= \left[\bm{v}(\bm{x}) - \bm{v}(\bm{x}')\right] \cdot \hat{z}
    = -i
    \int_{\bm{k}} e^{i \bm{k} \cdot (\bm{x} - \bm{x}') } \frac{\bm{k} \cdot \hat{z} }{k^2} \theta (\bm{k}) \notag \\
    &= -i\,aHf \int_{\bm{k}} e^{i \bm{k} \cdot (\bm{x} - \bm{x}') } \frac{\bm{k} \cdot \hat{z}}{k^2} \delta_L (\bm{k}) + {\cal O}\left[(\delta_L)^2\right] \notag \\
    &=
    \beta\,(-i\,aH) \int_{\bm{k}} e^{i \bm{k} \cdot (\bm{x} - \bm{x}') } \frac{\bm{k} \cdot \hat{z}}{k^2} \delta_{h} (\bm{k}) + {\cal O}\left[(\delta_L)^2\right] \ , \label{eqn:deltaur_beta}
\end{align}
where $\hat{z}$ stands for an arbitrary line of sight, and the integration symbol $\int_{\boldsymbol{k}}$ is a short-hand for $(2\pi)^{-3}\int{\rm d}^3\boldsymbol{k}$. In the second and third lines we have used that at linear order we can relate the velocity divergence to the linear matter perturbations as $\theta(\bm{k}) = aHf\,\delta_L(\bm{k})$, while the halo overdensity is given by $\delta_h(\bm{k}) = b_1\,\delta_L(\bm{k})$, showing that in linear perturbation theory $v_{\parallel}$ is determined by $\beta$ times a quantity depending only on the halo density field.

Consequently, at leading order the $n$-th moment of the pairwise velocity PDF,
\begin{equation}
    m_n(\bm{r}) = \frac{\left<v_{\parallel}^n \left[1 + \delta_h(\bm{x})\right]\,\left[1 + \delta_h(\bm{x}')\right]\right>}{1 + \xi_{hh}(r)}\,,
\end{equation}
where $\bm{r} = \bm{x}-\bm{x}'$, can be written as $\beta^n$ multiplied by a term depending on the halo auto power spectrum or correlation function. Explicitly, making use of Eq.~\eqref{eqn:deltaur_beta} and keeping only the leading order contributions, we obtain for the first moment:
\begin{align}
    [1+\xi_{hh}(r)]\,m_1(\bm{r}) &\approx 2 \left<v_{\parallel}\,\delta_h\right> \notag \\ &= 2 \beta\, aH\, \frac{\bm{r}\cdot \hat{z}}{r} \int_{\bm{k}} j_1(kr)\,\frac{P_{hh}(k)}{k}\,,
\end{align}
whereas the second moment gives
\begin{align}
    &[1+\xi_{hh}(r)]\,m_2(\bm{r}) \notag \\ 
    &\hspace{0.5em}\approx (1+\xi_{hh})\,\left<v_{\parallel}^2\right> + 2 \left<v_{\parallel}\,\delta_h\right>\left<v_{\parallel}\,\delta_h'\right> \notag \\ 
    &\hspace{0.5em}= 2 \beta^2\,(aH)^2 \left\{\left(\frac{\bm{r} \cdot \hat{z}}{r}\right)^2 \Bigg[ (1+\xi_{hh})\int_{\bm{k}} j_2(kr) \frac{P_{hh}(k)}{k^2} \right.\Bigg. \notag \\ &\hspace{12em}\Bigg.+\left(\int_{\bm{k}} j_1(kr) \frac{P_{hh}(k)}{k}\right)^2\Bigg] \notag \\
    &\hspace{1.5em}\Bigg.-\frac{1+\xi_{hh}}{3}\int_{\bm{k}}\big[j_2(kr)+j_0(kr)-1\big]\frac{P_{hh}(k)}{k^2}\Bigg\}\,,
\end{align}
and similar relations can be derived for the higher-order moments. 

Crucially, because in this work the halo catalogues from different gravity (or, more generally, different cosmological) models have been tuned so that they have the same halo correlation function $\xi_{hh}(r)$ and halo power spectrum $P_{hh}(k)$, these expressions show that, when taking ratios of the pairwise velocity moments from different models all terms involving $\xi_{hh}$ or $P_{hh}$ cancel. This leaves only factors of $\beta$, and given two models, $A$ and $B$, we therefore have
\begin{equation}
    \frac{m_{n,A}}{m_{n,B}} = \left(\frac{\beta_A}{\beta_B}\right)^n\,. \label{eqn:ratio_moments}
\end{equation}
It is important to stress that this only holds in linear theory and for that reason it is not guaranteed that Eq.~\eqref{eqn:ratio_moments} is valid on sufficiently large scales, as it is well known that for instance the large-scale variance receives significant contributions from small-scale virialised motions \citep{2004PhRvD..70h3007S}. For more discussion on this point and how this alters the ratio in Eq.~\eqref{eqn:ratio_moments} for even-order velocity moments, see Sec.~\ref{sec:halo_vel_moments}.

\section{Numerical details of the streaming model integration}
\label{appendix:details}

This appendix presents the numerical details in the computation of the streaming model predictions from the ingredients measured from simulations.
The streaming model for the redshift-space \ac{TPCF} (Eq.~\eqref{eqn:streaming_model_core}) has two ingredients: the real-space \ac{TPCF} and the line-of-sight pairwise velocity PDF, i.e. the position and velocity information of tracers.

As mentioned in Eq.~\eqref{eqn:method_ST_best_params_fitting}, we do not directly use $\mathcal{P} (v_{\parallel} | \bm{r})$ in our model predictions, but approximate it with the \ac{ST} distribution 
\[ 
    \mathcal{P}_{\rm ST} \big(v_{\parallel} | v_c(\bm{r}), w(\bm{r}), \alpha(\bm{r}), \nu(\bm{r}) \big).
\]
The four \ac{ST} parameters for a given pair separation $\bm{r}$ can be fixed by the first four line-of-sight pairwise veolcity moments $m_1$, $c_{2\text{-}4} (\bm{r})$, i.e., by solving the four nonlinearly coupled algebraic equations, Eqs.~(\ref{eqn:st_m1_moments}-\ref{eqn:st_g2_moments}).
This is done by using the \texttt{fsolve} function of the standard open-source \texttt{scipy} \citep{2020SciPy-NMeth} library.

The line-of-sight pairwise veolcity moments can be obtained by: 
\begin{itemize}
    \item either directly measuring the line-of-sight pairwise distribution $\mathcal{P} (v_{\parallel} | \bm{r})$, or 
    \item measuring the two-dimensional pairwise veolcity distribution $\mathcal{P} (v_r, v_t | r)$ and projecting its moments along the line of sight according to Eq.~\eqref{eq:moments_relation}.
\end{itemize}

\begin{figure}
    \centering
    \includegraphics[width=\columnwidth]{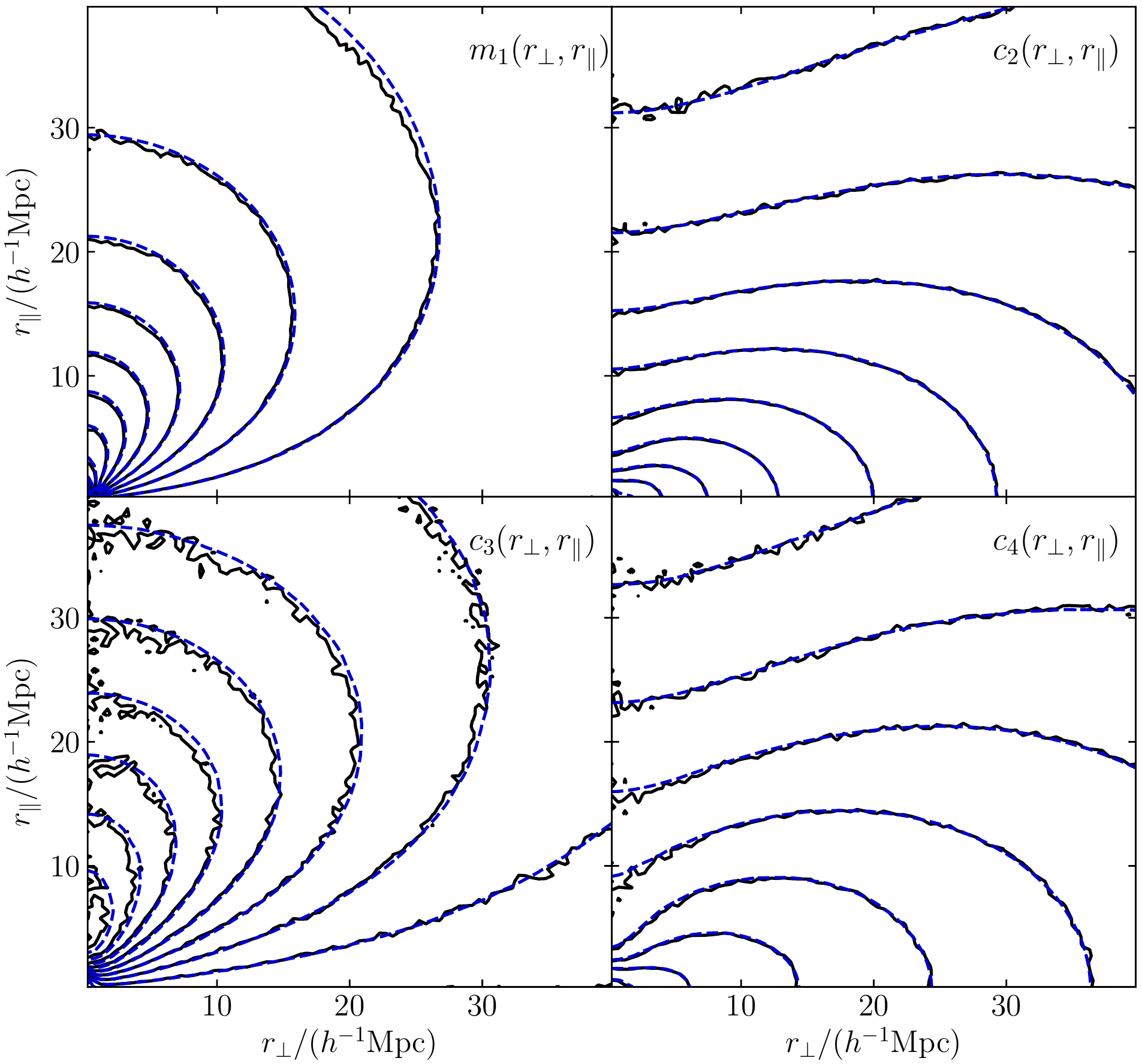}
    \caption{The first four orders of the line-of-sight halo pairwise velocity moments, from direct measurements (black solid lines) and pairwise velocity moments projection (blue dashed lines). The halo catalogues are from  the \textsc{glam} simulations with a fixed number density $n_h = 10^{-3.5} \, (h^{-1}\mathrm{Mpc})^{-3}$ at $z=0.5$.  }
    \label{fig:moments_project}
\end{figure}

We prefer the second approach since $\mathcal{P} (v_r, v_t | r)$ takes advantage of symmetries and does not require fixing a particular line-of-sight when using the simulation data.
We have checked that the projected line-of-sight moments from these two approaches are in good agreement. 
Fig.~\ref{fig:moments_project} shows the case of the halo catalogues from \textsc{glam} simulations with the number density $n_h = 10^{-3.5} \, (h^{-1}\mathrm{Mpc})^{-3}$ at $z=0.5$.
The \ac{ST} distribution with the model parameters obtained by this method has been compared with the measured line-of-sight velocity PDF in Fig.~\ref{fig:VlosPDF_MGGLAM_fRn1_NumDen3.0}.

In practice, the streaming model numerical integrals can be sensitive to various factors, such as the choice of integration method, the binning scheme etc., and it is important to make sure that one's choices lead to converged results. We have created an example code for this, which can be found \href{https://github.com/chzruan/StreamingModelExample}{here}.
The example code calculated the redshift-space correlation function multipoles for the \textsc{glam} halo catalogues with the number density $10^{-3.0} \, (h^{-1}\mathrm{Mpc})^{-3}$ at $z=0.25$. 
For the model ingredients $\xi^{\rm R}(r)$, $m_{10}(r)$ and $c_{(2)\text{-}(4)}(r)$, we measured them in the separation bins linearly spaced over $1 \le r / (h^{-1}\mathrm{Mpc}) \le 120$ with a bin width $\Delta r = 1\,h^{-1}\mathrm{Mpc}$.
The optimal configurations depend on the tracers' type (e.g. haloes versus galaxies), number density and redshift, etc.

\bsp	
\label{lastpage}
\end{document}